%% file: trantr_sc.tex
\documentclass[colorlinks,peerreview, draftcls, 12pt]{IEEEtran}

\usepackage{cite}

\ifCLASSINFOpdf
\else
\fi
\usepackage[cmex10]{amsmath} 
\usepackage{amsthm,amssymb}

\usepackage{array}

\usepackage{mdwmath}
\usepackage{mdwtab}

\usepackage[tight,footnotesize,it]{subfigure}

\input{myshort}

\begin{document}
%
\title{Robustness of Time Reversal vs. All-Rake Transceivers in Multiple Access Channels}

%
%
%

\author{Guido C. {Ferrante},~\IEEEmembership{Student Member,~IEEE,}
        Jocelyn {Fiorina},~\IEEEmembership{Member,~IEEE}, Maria-Gabriella {Di Benedetto},~\IEEEmembership{Senior Member,~IEEE}.%
\thanks{G.C. Ferrante is with the Department
of Information Engineering, Electronics and Telecommunications, Sapienza University of Rome,
Rome, ITALY, and with the Department of Telecommunications, Sup\'elec, Gif-sur-Yvette, FRANCE.}%
\thanks{J. Fiorina is with the Department of Telecommunications, Sup\'elec, Gif-sur-Yvette, FRANCE.}%
\thanks{M.-G. Di Benedetto is with the Department
of Information Engineering, Electronics and Telecommunications, Sapienza University of Rome, Rome, ITALY.}%
\thanks{This work has been submitted to the IEEE for possible publication. Copyright may be transferred without notice, after which this version may no longer be accessible.}}

%
%

\markboth{Submitted to the IEEE Transactions On Wireless Communications}{}%

%



\maketitle
\begin{abstract}
\boldmath
Time reversal, that is prefiltering of transmitted signals with time reversed channel impulse responses, may be used in single user communications in order to move complexity from the receiver to the transmitter, and in multiuser communications to also modify statistical properties of multiuser interference. Imperfect channel estimation may, however, affect pre- vs. post- filtering schemes in a different way. This issue is the object of this paper; Robustness of time reversal vs. All-Rake (AR) transceivers, in multiple access communications, with respect to channel estimation errors, is investigated. Results of performance analyses in terms of symbol error probability and spectral efficiency when the receiver is structured either by a bank of matched filters or by 1Rake, followed by independent decoders, indicate that AR is slightly more robust than time reversal but requires in practice more complex channel estimation procedures since all channels must be simultaneously inferred in the multiuser communication setting. 
\end{abstract}


%

\section{Introduction}
%
%
%
%

\IEEEPARstart{F}{ield} equivalence principles \cite{shelk:uno,shelk:due,stratton:uno,chen:math} state that the radiated field within a volume $V$, with boundary $\partial V$, enclosing a source, can be computed by considering, in place of actual source, an infinity of equivalent virtual sources placed on $\partial V$. 

Suppose the source $S_O$ is pointwise, impulsive, and located in a point $O\in V$. The electromagnetic problem of finding radiated field in $V$ can be solved based on the Green function.


From a communication perspective, knowing the channel at all points of $\partial V$ would allow, in principle, to understand the nature of a source $S_O$, that is, the location of a pointwise source within volume $V$, that radiated the field observed on $\partial V$. Sensing the channel on $\partial V$ would require a multiantenna system and a perfect knowledge of  impulse responses of channels between $O$ and all points on $\partial V$.

Time reversal is a technique that takes advantage of the above physical phenomenon and that was also proposed in acoustics \cite{fink:old,derode:aps,fink:acoustic}. By prefiltering transmissions with a scaled version of the channel impulse response, reversed in time, allows simplification of receiver design, since the channel is compensated by precoding. Time reversal also focuses signals in space, given that there is only one ``correct'' location of the receiver that experiences the specific ``time reversed'' channel impulse response.

Pioneering work on single-antenna time reversal spread-spectrum communications dates back to the nineties, where the time reversal pre-filter was named \textit{pre-Rake} \cite{esma:rake1,esma:rake2}. The basic idea was to pre-filter the transmitted pulse with the channel impulse response reversed in time, therefore matching the transmitted signal with the subsequent channel. 

Precoding techniques for multiuser spread-spectrum systems were developed along similar lines of receive filters: transmit Zero-Forcing (ZF) \cite{tang:zfds}, that attempts to pre-equalize the channel by flattening the effective channel formed by the cascade of the pre-filter and the actual channel, is optimum in the high-$\SNR$ regime; transmit matched-filter (MF), that has been recognized to be equivalent to the pre-Rake filter in \cite{joham:equiv}, that conversely is optimum in the low-$\SNR$ regime; and finally, transmit MMSE (Wiener) filter minimizing the $\SINR$ was derived in \cite{joham:mmse} following previous attempts \cite{voj:tx,noll:ss}.

In recent years, along with the fast developing of narrowband MIMO systems, pre-coding techniques using multiple antennas at the transmitter were thoroughly studied (for a complete overview on MIMO precoding see \cite{precoding:overview}). Since the mathematical formulation of multiuser spread-spectrum is very close to that of MIMO communications (see \cite{palo:mono} for an overview of this analogy), MIMO linear precoders can be derived along similar techniques.

Time reversal was proposed in connection to UWB communications in \cite{stro:tr}, that also addressed equalization through an MMSE receiver. In \cite{qiu:symp}, early experimental data, showing the feasibility of time reversal, were collected. Following, experimental investigations on multiple-antenna systems with time reversal \cite{ngu:trmimo,qiu:misoexp1,qiu:mimo}, and performance analyses \cite{qiu:misoexp2}, were also pursued. In \cite{qiu:gen1,qiu:gen2}, compensation for pulse distortion in connection to time reversal was investigated. In \cite{ferr:wcnc}, the trade-off between the complexity of transmitter vs. receiver in terms of number of paths was analyzed. In \cite{ferr:icuwb13}, the insensitivity of time reversal to the lack of correlation between channels in a MISO system was investigated. Finally, in \cite{denardis:uwbtr}, the effect of time reversal on statistical properties of multiuser interference in communication vs. positioning was explored.

The above investigations were all carried out based on the hypothesis of perfect channel estimation. This hypothesis, however, is strong, since it is unrealistic, irrespectively of whether channel estimation is performed at the transmitter or at the receiver. While previous papers addressed the comparison of pre- and post- channel filtering, the problem remains of realistic imperfect channel estimation and of how this affects performance for time reversed vs. receiver-based channel estimation schemes. 

This paper addresses the above problem, by comparing single-antenna systems using time reversal, against receiver-based equalization schemes such as the exemplary case of an {AR} receiver. 

The adopted network model considers multiple access by $K$ user terminals (UTs) communicating to one basestation (BS), where both UTs and the BS have one antenna only, and  communication between each UT and BS adopts ultra-wideband, impulse-radio signaling.

Comparison of performance of time reversal vs. {AR} transceivers will be carried out in terms of effect of imperfect channel state information (CSI) on symbol error probability of a generic information-bearing symbol for a given UT (see \cite{sens:cdma} for a work on a close topic regarding CDMA systems). The analysis will further explore robustness of time reversal vs. {AR}, by finding the maximum achievable rate for the uplink channel. Finally, the maximum information rate, that takes into account channel estimation overhead, will be explored. 

The paper is organized as follows: Section~\ref{sec:model} contains the system model; 
Section~\ref{sec:poe} is devoted to the performance analysis in terms of symbol error probability; Section~\ref{sec:arates} contains results of comparison in terms of uplink rate of the network. Section~\ref{sec:conc} contains the conclusions.


\section{Reference Model}\label{sec:model}

This section is organized as follows: \ref{sub:nmodel}. \textit{Network model}; \ref{sub:suc}. \textit{Single User Channel}; \textit{Multiuser Channel}; \ref{sub:cedt}. \textit{Channel Estimation and Data Transmission}. 

\subsection{Network Model}\label{sub:nmodel}
\begin{figure}[tb]
\vspace{1cm}
\centering

%
%
%
%
%
\includegraphics{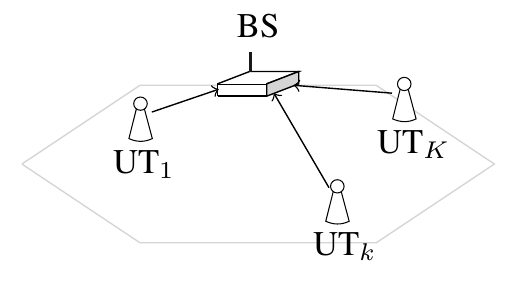}
\caption{Network model: several user terminals (UTs) transmit information-bearing symbols to a common sink, \ie, basestation (BS).}
\label{fig:uplink1}
\end{figure}

A multiple access channel where $K$ independent sources transmit information-bearing symbols to a common sink is considered (uplink communication channel). Borrowing the terminology from the cellular network field, sources of information are called user terminals (UTs) and the sink is called basestation (BS). However, UTs and BS are intendend to designate more than what the name implies. For example, in a typical WLAN, a BS is a fixed (\eg desktop) or mobile receiver (\eg tablet, laptop, mobile phone) and UTs are peripherals or other fixed vs. mobile devices. Figure~\ref{fig:uplink1} shows the adopted network model.

A generic $\UT_k$ transmits data, that is encoded into a sequence of information-bearing symbols $\{b_k[m]\colon m\in\Z\}$. This set of symbols is partitioned into blocks of $n$ symbols each, $\{ \bs{b}_k[i]\colon i\in\Z \}$ where $\bs{b}_k[i]=(b_k[in],\dotsc, b_k[(i+1)n-1])^\t$. The length $n$ of the block in terms of symbols will be linked below with the coherence time $T_\mrm{coh}$ of the channel. Each block is transmitted by the following signal:
\begin{equation}\label{eq:txgen} s_{k,i}(t;\bs{b}_k[i]) \hspace{2mm}=\! \sum_{m=in}^{(i+1)n-1} b_k[m] \gT{k,m}(t-mT_\msf{s}), \end{equation}
where $T_\msf{s}$ (sec) is the symbol period and $\gT{k,m}(t)$ is the unit energy waveform associated with the $m$-th symbol of user $k$. In general, $\gT{k,m}(t)$ is a spread-spectrum signal at user $k$ prefilter output, and has band $[-\msf{W}/2,\msf{W}/2]$, that is, its spectrum is nonzero for $|f|\leq \msf{W}/2$. Assuming that $\{\gT{k,m}(t-mT_\msf{s})\colon m=in,\dotsc,(i+1)n-1\}$ are orthonormal, or very mildly crosscorrelated, the energy of $s_{k,i}(t;\bs{b}_k[i])$ in eq.~\eqref{eq:txgen} is $n\E{|b_k[m]|^2}$; Since the block has duration $nT_\msf{s}$, the average power is $\mathcal{P}_{\!k}=\E{|b_k[m]|^2}/T_\msf{s}$. 

\begin{figure}[t]
\centering 
\includegraphics[scale=0.9]{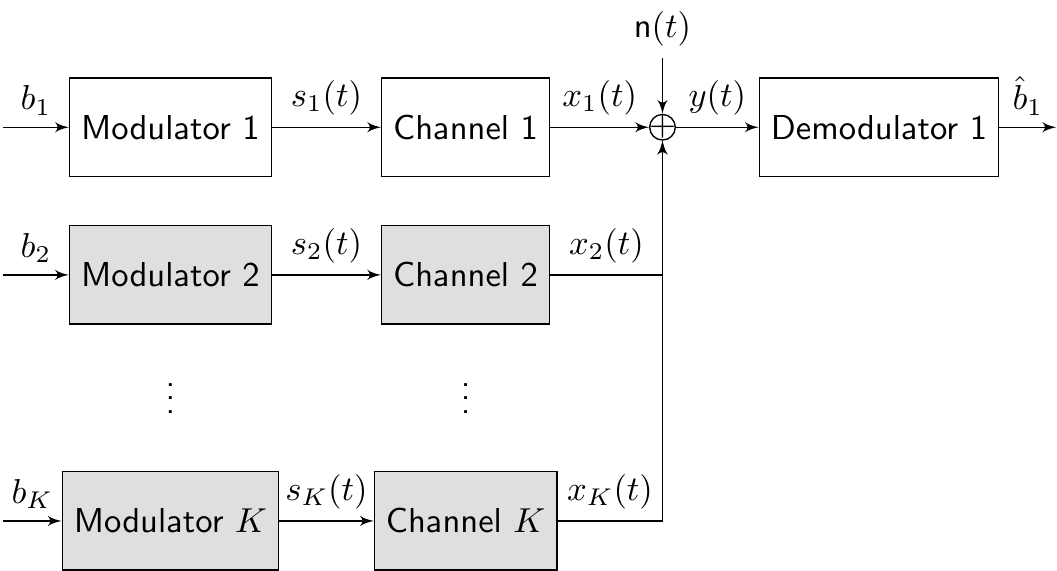}
\caption[]{System model. The transmitter is formed by $K$ modulators. Radiated signals are affeted by $K$ different channels and white gaussian noise $n(t)$ at the receiver. Receveir consists in one demodulator shown on figure for the example case of user $1$.}
\label{fig:uplink2} 
\end{figure}

In the adopted model, demodulation at BS is performed on a block-by-block basis. Index $i$, that specifies the block number, is thus dropped. Consider for the sake of simplicity $i=0$ in eq.~\eqref{eq:txgen}:
\begin{equation}\label{eq:seqtx} s_{k}(t;\bs{b}_k) = \sum_{m=0}^{n-1} b_k[m] \gT{k,m}(t-mT_s). \end{equation}
Figure~\ref{fig:uplink2} shows the system model, including $K$ modulators producing $K$ transmitted signals, $s_k(t)$, $k=1,\dotsc,K$, affected by propagation within $K$ different channels and corrupted at the receiver by white gaussian noise $n(t)$. The receiver consists in one demodulator.

Transmitted signal $s_k(t;\bs{b}_k)$ of each user propagates over a multipath channel with impulse response $c_k(t)$ and is distorted into $x_k(t)$:
\begin{equation}\label{eq:rxdist} x_k(t;\bs{b}_k) = s_k(t;\bs{b}_k) * c_k(t) = \sum_{\ell=0}^{\infty} c_{k,\ell} s_k(t-\tau_{k,\ell};\bs{b}_k), \end{equation}
where $\{c_{k,\ell}\colon \ell\geq 0\}$ and $\{\tau_{k,\ell}\colon \ell\geq 0\}$ are amplitudes and delays of the paths of $c_k(t)$, respectively. 

The received signal is:
\begin{equation}\label{eq:rxct}  y(t;\bs{b}) = \sum_{k=1}^K x_k(t;\bs{b}_k) + n(t), \end{equation}
where $n(t)$ is a white Gaussian noise with flat power spectral density $\N/2$ (W/Hz). %
Throughout the paper, the receiver estimates transmitted symbols of user $k$, $\{ b_k[m] \colon k=1,\dotsc,K; m=0,\dotsc,n-1 \}$, on a symbol-by-symbol basis, by considering users $j\neq k$ as unknown interference over user $k$; for example, Fig.~\ref{fig:uplink2} shows the demodulation of user $1$. As detailed below, transmissions are symbol-synchronous {but not necessarily chip-synchronous}, therefore the symbol-by-symbol demodulation does not imply any performance loss. In the adopted model, the receiver is a single user detector, and as such suboptimal, since it does not take into account the possibility of joint multiuser detection. How channel is estimated and how error affected estimated channels play a role in the model will be explained further down in this section in association with the different modulation and demodulation structures. %
Expliciting signals for the symbol at time epoch $m=0$, and denoting by $b_k=b_k[0]$, eqs.~\eqref{eq:seqtx},~\eqref{eq:rxdist}~and~\eqref{eq:rxct} become:
\begin{align}
\label{eq:seqtxONE} s_{k}(t;{b}_k) & = b_k \gT{k,0}(t), \\ 
 x_k(t;{b}_k) & = s_k(t;{b}_k) * c_k(t) = \sum_{\ell=0}^{\infty} c_{k,\ell} s_k(t-\tau_{k,\ell};{b}_k), \label{eq:rxdistONE} \\[-3mm]
\label{eq:rxctONE}  y(t;{b_1,\dotsc,b_K}) & = \sum_{k=1}^K x_k(t;{b}_k) + n(t). 
\end{align}

\subsection{Single User Channel}\label{sub:suc}
Since the system symbol-synchronous, analysis may refer to transmission of one generic symbol, that is chosen as symbol $m=0$, $b[0]$, denoted by $b$. If transmission does not foresee prefiltering, that is, a zero-excess bandwidth pulse $\psi(t)$ with bandwidth $\msf{W}$ and unit energy is transmitted to modulate $b$, the received signal is:
\begin{align}\label{eq:rxcompleteonebit} y(t;b) = x(t;b)+n(t) 
		 & \overset{\textup{(a)}}{=} b \sum_{\nu=0}^{N-1} \msf{s}[\nu] \psi(t-\nu T_\msf{c})*c(t) + n(t) \nonumber \\
 		 & \overset{\textup{(b)}}{=} b \sum_{\nu=0}^{N-1} \msf{s}[\nu] \sum_{\ell=0}^{L\ii} c[\ell]\psi(t-\ell/\msf{W}-\nu T_\msf{c}) + n(t) \nonumber \\
 		 & \overset{\textup{(c)}}{=} b \sum_{\nu=0}^{N-1} \msf{s}[\nu] \sum_{\ell=0}^{L\ii} c[\ell]\psi(t-(\ell+\nu\ii)/\msf{W}) + n(t),
\end{align}
where in $\textup{(a)}$ the spreading sequence $\bs{s}=(\msf{s}[0],\dotsc,\msf{s}[N-1])^\t$ and the chip period $T_\msf{c}$ are made explicit; $\textup{(b)}$ follows from $\psi(t)$, and therefore also $\psi(t)*c(t)$, being bandlimited to $\msf{W}/2$; and $\textup{(c)}$ follows from assuming $T_\msf{c}=\ii/\msf{W}$, being $\ii$ a positive integer called \textit{impulsiveness index}, that introduces a model to account for a pulse duration shorter than chip duration, as common in UWB communications. In the following, time-hopping is considered, for which all $\msf{s}[\nu]$ are zero, but one. 

By projecting eq.~\eqref{eq:rxcompleteonebit} onto $\{ \psi(t-k/\msf{W})\colon k=0,\dotsc,(N+L+1)\ii-1-1 \}$, the following discrete model is obtained:
\begin{equation}\label{eq:discrSUonebit} \bs{y} = \bs{C} \bs{x}\, b+\bs{n},\quad \bs{x}=\bs{s}\otimes \bs{e}^\ii_1, \end{equation}
where $\bs{e}^\ii_1=[1, \bs{0}^\t_{(\ii-1)\times 1}]^\t$ is the first vector of the canonical basis of $\R^\ii$, $\bs{x}$ is $N\ii \times 1$, and $\bs{C}$ is Toeplitz with dimensions $(N+L+1)\ii-1 \times N\ii$ and elements $C_{ij}=c[i-j]$. Since $c(t)$ is causal, it is assumed that $c[\ell]=0$ for $\ell<0$, and since $c(t)$ has finite delay spread $T_d$, it is assumed that $c[\ell]=0$ for $\ell>L\ii$, being $T_d=LT_\msf{c}=L\ii/\msf{W}$; therefore, $\bs{C}$ is banded Toeplitz. 

\medskip

In general, for a system with prefiltering, with prefiltering impulse response $p(t)$, eq.~\eqref{eq:discrSUonebit} generalizes to (see \eg \cite{palo:mono}):
\begin{equation}\label{eq:discrSUonebitPRE} \bs{y} = \bs{C} \bs{P} \bs{x}\, b+\bs{n}, \end{equation}
where $\bs{P}$ is a Toeplitz matrix with dimensions $(N+2L)\ii\times(N+L+1)\ii-1$ and elements $P_{ij}=p[i-j]=p*\psi((i-j)/\msf{W})$. 

In this paper, prefiltering is introduced in order to compensate channel effects; in particular, prefiltering is based on an estimated version of the channel impulse response. In other words, imperfect prefiltering may be matched to channel estimation error patterns. If prefiltering is imperfect, as will be justified in Subsection~\ref{sub:cedt}, the error due to the estimation process can be modeled as a white Gaussian process $\xi(t)$, that is added to $\bs{P}$ as follows:
\begin{equation}\label{eq:pimperf}
\bs{\hat{P}} = \alpha(\bs{P}+\bs{\Xi}),
\end{equation}
where $\Xi_{ij} = \xi_{i-j}\sim\mathcal{N}(0,\sigma_\xi^2)$, where $\sigma_\xi^2$ accounts for estimation accuracy, and $\alpha>0$ is such that $\| \bs{\hat{P}}\bs{x} \|^2=\| \bs{P}\bs{x} \|^2$.


\bigskip
\noindent\textit{{No prefiltering, All-Rake receiver}. } 

The traditional (or conventional) receiver is a matched-filter, \ie, an {AR} receiver in the case of a multipath channel. Knowing the time-hopping spreading sequence $\bs{x}$ and the resolved channel $\bs{c}$, a sufficient statistic for $b$ is obtained by projecting the received signal $\bs{y}$ onto $\bs{h}=\bs{C}\bs{x}$, or, equivalently, onto $\bs{h}/\|\bs{h}\|$:
\[ z^\mrm{AR}=\frac{\bs{h}^\t}{\|\bs{h}\|} \bs{y} = \|\bs{h}\| b + \nu, \]
where $\nu\sim\mathcal{N}(0,\N/2)$. 

As occurs in the prefiltering, also the {AR} receiver is affected by possible channel estimation errors. If the {AR} is provided with imperfect channel state information (CSI), that is, operates using an estimation $\bs{\hat{c}}$ of channel $\bs{c}$ that is impaired by an error $\bs{\xi}\sim\mathcal{N}(\bs{0},\sigma_\xi^2\bs{I}_{(L\ii+1)})$, then the {AR} combines paths through $\bs{\hat{h}}\eqdef \bs{\hat{C}}\bs{x}$ instead of $\bs{h}=\bs{C}\bs{x}$, and inference of $b$ is based on:
\begin{align}\label{eq:decR} 
\hat{z}^\mrm{AR} = \frac{\bs{\hat{h}}^\t}{\|\bs{\hat{h}}\|} \bs{y} 
	 = \frac{(\bs{h}+\bs{\chi})^\t}{\|\bs{h}+\bs{\chi}\|}(\bs{h}\.b+\bs{n})
	 = \frac{(\bs{h}+\bs{\chi})^\t}{\|\bs{h}+\bs{\chi}\|}\bs{h}\.b+\frac{(\bs{h}+\bs{\chi})^\t}{\|\bs{h}+\bs{\chi}\|}\bs{n},
\end{align}
where $\bs{\chi}=[\bs{0}_{j_{\bs{x}}}^{\t},\, \bs{\xi}^\t,\, \bs{0}_{N\ii-j_{\bs{x}}}^\t]^{\t}$, being $j_{\bs{x}}$ the nonzero dimension of $\bs{x}$.

\bigskip
\noindent\textit{{Time Reversal prefiltering, 1Rake receiver}. } 

The time reversal prefilter is represented by $p[j]=\alpha c[L\ii-j]$, where $\alpha>0$ guarantees that prefiltered and non prefiltered transmitted waveforms have same energy. Time-hopping implies $\bs{s}\in\{\bs{e}_\nu\}_{\nu=1}^N$, and $\bs{x}=\bs{s}\otimes \bs{e}^\ii_1\in\{\bs{e}_{(\nu-1)\ii+1}\}_{\nu=1}^N$. A 1Rake receiver is given by $\bs{e}_{L\ii+j_{\bs{x}}}$. Denoting by $\bs{T}$ the time-reversal prefilter matrix, one has:
\begin{align}\label{eq:zTRperfect}
 z^\mrm{TR}			 = \bs{e}_{L\ii+j_{\bs{x}}}^\t \bs{y}
 						 = \bs{e}_{L\ii+j_{\bs{x}}}^\t \bs{C} \bs{T} \bs{x}\, b+\bs{e}_{L\ii+j_{\bs{x}}}^\t \bs{n}
						 = \bs{e}_{L\ii+j_{\bs{x}}}^\t \bs{H} \bs{x} \, b + n_{L\ii+j_{\bs{x}}},
\end{align}
being $n_{L\ii+j_{\bs{x}}}\sim\mathcal{N}(0,\N/2)$.

If the transmitter is provided with imperfect CSI, then model of eq.~\eqref{eq:pimperf} holds, and eq.~\eqref{eq:zTRperfect} becomes:
\begin{align}\label{eq:zTR}
 \hat{z}^\mrm{TR} 	 = \bs{e}_{L\ii+j_{\bs{x}}}^\t \bs{C} \bs{\hat{T}} \bs{x}\, b+\bs{e}_{L\ii+j_{\bs{x}}}^\t \bs{n} 
			 = \bs{e}_{L\ii+j_{\bs{x}}}^\t \bs{C} [\alpha(\bs{T}+\bs{\Xi})] \bs{x}\, b+ n_{L\ii+j_{\bs{x}}}.
\end{align}

\bigskip
\noindent\textit{{AR} vs. {TR}. } 

As well-known \cite{denardis:uwbtr}, {TR} is equivalent to a system without prefiltering and {AR} in terms of the signal-to-noise ratio. From a single user perspective, there is no performance difference in both uncoded (symbol error probability) and coded (channel capacity) regimes between the two transceiver structures. Moreover, previous work \cite{ferr:wcnc} suggested that sets of equivalent systems can be obtained with partial Rakes compensating for partial time reversal transmitter structures. In the case of imperfect CSI, the comparison of the different structures is the object of this paper. 

\subsection{Multiuser Channel}\label{sub:mc}

A straightforward extension of eq.~\eqref{eq:discrSUonebit} to $K$ users is as follows:
\begin{equation}\label{eq:discrMUonebit} \bs{y} = \sum_{k=1}^K \bs{C}_{k} \bs{x}_k\, b_k+\bs{n}, \end{equation}
where $\bs{x}_k=\bs{s}_k\otimes \bs{e}^\ii_{l_k}$ and $1\leq l_k \leq \ii$ models the chip-asynchronism by making $l_k$ i.i.d. according to a uniform distribution. This extension holds based on the hypothesis that all UTs are symbol-synchronous. This hypothesis is reasonable since, as further discussed in Subsection~\ref{sub:cedt}, the BS broadcasts in a link setup phase a known sequence to the UTs. %
Denoting by $\bs{h}_k=\bs{C}_{k} \bs{x}_k$ the spreading sequence $\bs{x}_k$ after transition in the multipath channel, and by:
\[ \bs{H}=\sum_{k=1}^K \bs{e}_k^\t \otimes \bs{h}_k = [\bs{h}_1,\dotsc,\bs{h}_K] \]
the spreading matrix, eq.~\eqref{eq:discrSUonebit} can also be rewritten as follows:
\begin{equation}\label{eq:discrMUGEN} \bs{y} = \bs{H}\bs{b}+\bs{n}, \end{equation}
where $\bs{b}=(b_1,\dotsc,b_K)^\t$. %
For systems with prefiltering, eq.~\eqref{eq:discrMUonebit} generalizes to:
\begin{equation}\label{eq:discrMUonebitPRE} \bs{y} = \sum_{k=1}^K \bs{C}_{k}\bs{P}_k \bs{x}_k\, b_k+\bs{n},\end{equation}
where matrices $\bs{P}_k$ and $\bs{C}_k$ have same dimensions as $\bs{P}$ and $\bs{C}$ of eq.~\eqref{eq:discrSUonebitPRE}, respectively, and eq.~\eqref{eq:discrMUGEN} holds with $\bs{h}_k=\bs{C}_{k} \bs{P}_k \bs{x}_k$. %
In the presence of imperfect CSI, $\bs{P}_k$ in eq.~\eqref{eq:discrMUonebitPRE} is substituted by $\bs{\hat{P}}_k$, as defined in eq.~\eqref{eq:pimperf}, where estimation errors are independent with respect to $k$.

\bigskip
\noindent\textit{{No prefiltering, All-Rake receiver}. }

The decision variable following the matched filter of user $k$ is:
\begin{align}\label{eq:zARperfect}
z^\mrm{AR}_{k} 
	 = \|\bs{h}_k\| b_k + \sum_{\substack{j=1\\j\neq k}}^K \frac{\bs{h}_k^\t}{\|\bs{h}_k\|} \bs{h}_j\, b_j + \frac{\bs{h}_k^\t}{\|\bs{h}_k\|} \bs{n}
	 = \|\bs{h}_k\| b_k + I_{k} + \nu_k,
\end{align}
where $\bs{h}_i=\bs{C}_{i} \bs{x}_i$, $I_{k}$ represents the MUI, and $\nu_k=\frac{\bs{h}_k^\t}{\|\bs{h}\|} \bs{n}\sim\mathcal{N}(0,\N/2)$.

If the {AR} is provided with imperfect CSI, then signal $\bs{y}$ is projected onto $\bs{\hat{h}}_k$ instead of $\bs{h}_k$, hence: 
\begin{align}\label{eq:zARimperfect}
\hat{z}^\mrm{AR}_{k} 
	 = \frac{\bs{\hat{h}}_k^\t}{\|\bs{\hat{h}}_k\|}\bs{h}_k b_k + \sum_{\substack{j=1\\j\neq k}}^K \frac{\bs{\hat{h}}_k^\t}{\|\bs{\hat{h}}_k\|} \bs{h}_j\, b_j + \frac{\bs{\hat{h}}_k^\t}{\|\bs{\hat{h}}_k\|} \bs{n} 
	 = \hat{a}_{kk}^\mrm{AR} b_k + \hat{I}_{k}^\mrm{AR} + \hat{\nu}_k^\mrm{AR},
\end{align}
where $\hat{\nu}_k^\mrm{AR}=\frac{\bs{\hat{h}}_k^\t}{\|\bs{\hat{h}}_k\|} \bs{n}\sim\mathcal{N}(0,\N/2)$. 

\bigskip
\noindent\textit{{Time Reversal prefiltering, 1Rake receiver}. }

With time reversal, the decision variable for user $k$ becomes:
\begin{equation}\label{eq:zTRperfectMU}
 z^\mrm{TR}_k = \bs{e}_{q_k}^\t \bs{C}_k \bs{T}_{k} \bs{x}_k\, b_k +\sum_{\substack{j=1\\j\neq k}}^K  \bs{e}_{q_k}^\t \bs{C}_j \bs{T}_{j} \bs{x}_j\, b_j + \bs{e}_{q_k}^\t \bs{n},
\end{equation}
where $q_k=L\ii+j_{\bs{x}_k}$ is the delay (in samples) to which the 1Rake is synchronized. In the presence of imperfect CSI, the decision variable is:
\begin{equation}\label{eq:zTRimperfect}
 \hat{z}^{\mrm{TR}}_{k}
 	 =  \bs{e}_{q_k}^\t \bs{C}_k \alpha_k (\bs{T}_{k}+\bs{\Xi}_k) \bs{x}_k\, b_k + 
	 	+ \sum_{\substack{j=1\\j\neq k}}^K \alpha_j \bs{e}_{q_k}^\t \bs{C}_j  (\bs{T}_{j} +\bs{\Xi}_j) \bs{x}_j\, b_j 
		+ \bs{e}_{q_k}^\t \bs{n}.
\end{equation}

\bigskip
\noindent\textit{{AR} vs. {TR}. }

As well-known \cite{denardis:uwbtr,fiorina:icuwb11}, time reversal usually increases the kurtosis of the interference at the output of Rake receivers. This follows from the fact that the effective channel impulse response formed by the combination of prefilter and multipath channel has a peaked behavior, whereas without time reversal the behavior is non-peaked. While in the single user case the two schemes are equivalent, this equivalence does not hold in the multiuser case. The impact of {estimation errors will be investigated below. }

\subsection{Channel Estimation and Data Transmission}\label{sub:cedt}
For both {AR} and {TR}, the channel impulse response estimation takes place, at least partially, both in the transmitter and in the receiver.

\begin{figure}[t]
\centering
\includegraphics{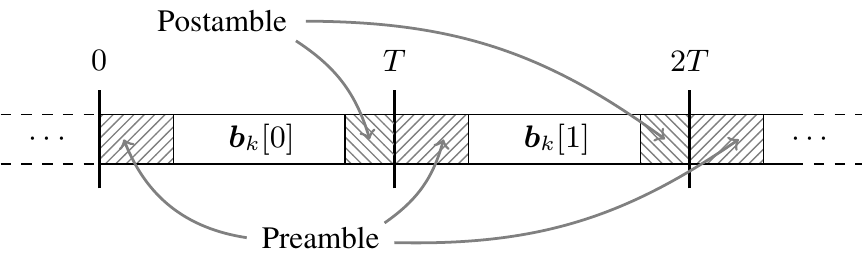}
\caption{Data transmission structured into blocks for $\UT_k$.}
\label{fig:block}
\end{figure}

Actual transmission of the set of information-bearing symbols requires, therefore, additional symbols to be sent either in a preamble or in a postamble of the block \cite{Larsson:mimo}, as shown in Fig.~\ref{fig:block}. 

Training is the simplest estimation process to evaluate the necessary channel state information. Time-Division Duplexing (TDD) is assumed as commonly witnessed in impulse-radio as well as common WLAN transmissions. Since precoding of $\UT_k$ does not depend on channels experienced by users $j\neq k$, feedback is not a necessary feature, given that channel is reciprocal. Note that there is no dedicated training since precoding is supposed to be disjoint, that is, the precoding vector of each UT does uniquely depend on the channel between its transmitter and the BS and is in particular independent of channels and precoding vectors of other UTs (see \cite{caire:bestpaper} for a thorough discussion). 

\begin{figure}[t]
\centering
\setlength\ploth{0.7\columnwidth} 
\setlength\plotw{0.875\columnwidth}
\subfigure[\emph{}]{\label{fig:txschemeglance}  \includegraphics[scale=0.925]{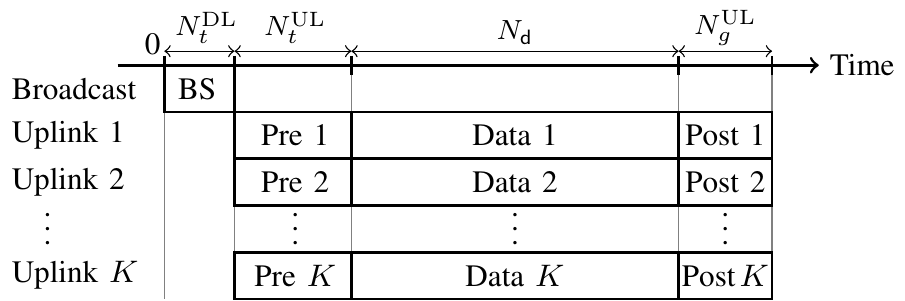} }
\subfigure[\emph{ }]{ \label{fig:txschemedetailed}\includegraphics[scale=0.9]{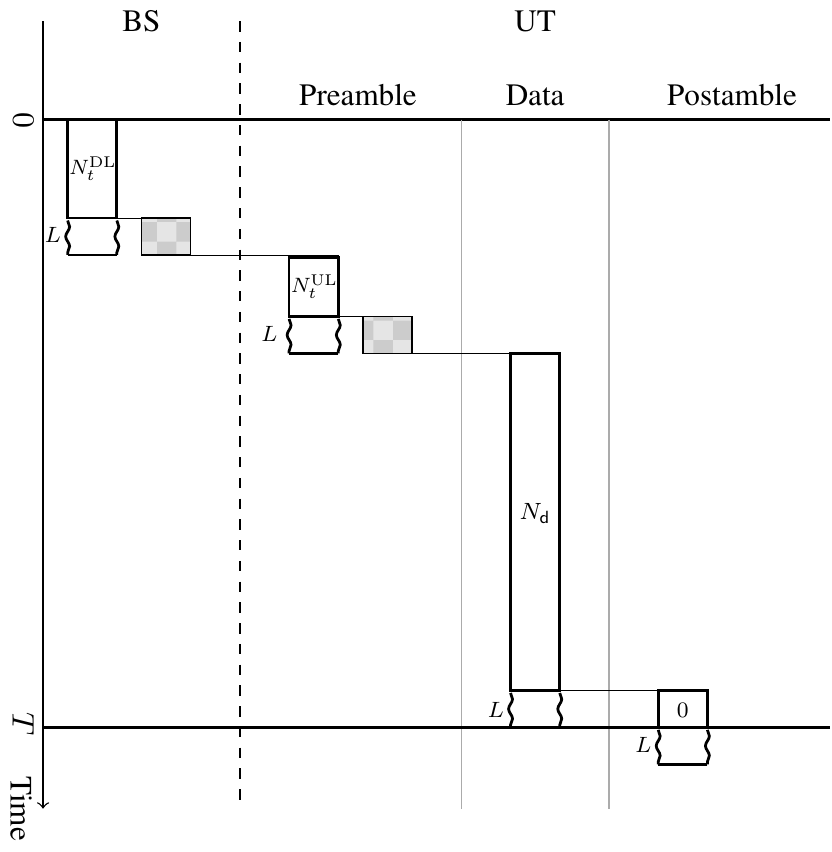} }
\caption{Transmission schemes. In \textit{(a)}: transmission scheme at a glance. In \textit{(b)}: detailed transmission scheme. \textit{Phase 1}: the BS broadcasts a training sequence of length $N_t^\mrm{DL}\ii$ samples (corresponding to $N_t^\mrm{DL}\ii/\msf{W}$ seconds) that is received by each UT starting at time $0$. Each multipath channel spreads the sequence for $L\ii$ samples, hence each UT listens from time $0$ to time $N_t^\mrm{DL}\ii+L\ii$ samples. This training sequence may be also used for network synchronization at symbol level. \textit{Phase 2}: Once the training sequence is received, each UT transmits its own training sequence of length $N_t^\mrm{UL}\ii$ samples to the BS (preamble). By reciprocity, channels spread these sequences for $L\ii$ samples, therefore each UT remains idle for $L\ii$ samples. \textit{Phase 3}: Each UT transmits a sequence of information-bearing symbols for $N_\msf{d}\ii=nN\ii$ samples. \textit{Phase 4}: Each UT transmits a sequence of null symbols denoted with $0$ (postamble).}
\label{fig:txschemes}
\end{figure}

Transmission follows a scheme that is shown at a glance in Fig.~\ref{fig:txschemeglance} and in more detail in Fig.~\ref{fig:txschemedetailed}. Figure~\ref{fig:txschemes} summarizes the organization of the different links (downlink, \ie, broadcast, vs. uplinks) over time, where durations of data, preamble and postamble are indicated in terms of number of chips ($N_t^\mrm{DL}$, $N_t^\mrm{UL}$, $N_\msf{d}$, $N_g^\mrm{UL}$). In particular, Fig.~\ref{fig:txschemedetailed} shows an information exchange between the BS and each UT consisting in four phases:
\begin{enumerate}
\item \textbf{Downlink Channel Training}: the BS broadcasts a training sequence of length $N_t^\mrm{DL}\ii$ samples, known by the set of UTs, followed by a zero-padding sequence of $L\ii$ samples (idle period), that allows each UT to receive the training sequence smeared by the channel; each UT estimates the channel based on the received samples and the knowledge of the transmitted training sequence, as further detailed in the remaining part of this section. This training sequence may be also used for network synchronization at symbol level.
\item \textbf{Uplink Channel Training}: each UT transmits a training sequence of length $N_t^\mrm{UL}\ii$ samples, known by the BS, followed by a zero-padding sequence of length $L\ii$ samples; these training sequences are assumed pseudo-noise (PN) sequences rather than orthogonal given that each UT chooses its sequence independently from the others. The BS estimates channels through the observation of the superposition of training sequences, that have been distorted by respective channels.
\item \textbf{Data Transmission}: each UT sends $n$ information-bearing symbols of a block, corresponding to $N_\msf{d}\ii=nN\ii$ samples; 
\item \textbf{Idle}: each UT sends a zero-padded postamble of duration $L\ii$ samples.
\end{enumerate}



\bigskip

During the downlink training, the BS broadcasts its training sequence to the UTs. With reference to model of Section~\ref{sec:model}, and in particular to eq.~\eqref{eq:discrSUonebit} and impulsiveness index $\ii$, the received signal at $\UT_k$ is:
\begin{equation}\label{eq:dltraining} \bs{y}^\mrm{DL}_k = \bs{C}_{k} \bs{\ups}^\mrm{DL}\!+\bs{n}_k,\quad \bs{\ups}^\mrm{DL}=\bs{\phi}_k^\mrm{DL} \otimes \bs{e}^\ii_1, \end{equation}
where $\bs{y}^\mrm{DL}_k$ is the $(N_t^\mrm{DL}\ii+L\ii)\times 1$ vector of received samples, $\bs{C}_{k}$ is the $(N_t^\mrm{DL}\ii+L\ii)\times N_t^\mrm{DL}\ii$ Toeplitz channel matrix, $\bs{\phi}_k^\mrm{DL}$ is the $N_t^\mrm{DL}\times 1$ training sequence, $\bs{\ups}^\mrm{DL}$ is the $N_t^\mrm{DL}\ii\times 1$ training sequence accounting for impulsiveness, and $\bs{n}_k$ is the $(N_t^\mrm{DL}\ii+L\ii)\times 1$ white Gaussian noise vector. 

Eq.~\eqref{eq:dltraining} can be rewritten as follows:
\begin{equation}\label{eq:dltrainingbis} \bs{y}^\mrm{DL}_k = \bs{\Ups}^\mrm{DL} \bs{c}_k+\bs{n}_k, \end{equation}
where now $\bs{\Ups}^\mrm{DL}$ is a $(N_t^\mrm{DL}\ii+L\ii)\times L\ii$ Toeplitz matrix and $\bs{c}_k$ is the $L\ii\times 1$ channel vector.

In order to minimize the signal-plus-interference-to-noise ratio, $\UT_k$ may use an MMSE estimation of $\bs{c}_k$, where the cause of interference is due to multipath. However, the use of PN sequences as training sequences is very common, due to their good autocorrelation properties. In fact, PN sequences have periodic ACF of the following form \cite{sarwate:cross,mitra:cross}:
\[ \rho_\phi[i]=\|\bs{\phi}\|^2\left(-\frac{1}{N_t^\mrm{DL}} + \delta_{i,0}\right), \quad 0 \leq i \leq N_t^\mrm{DL}-1, \]
that is asymptotically impulse-like, as $N_t^\mrm{DL}\gg 1$.

Asymptotically then, and dropping the superscript $\mrm{DL}$ to unclutter notation, $\bs{\Ups}^\t \bs{\Ups}\approx \|\bs{\phi}\|^2 \bs{I}= \|\bs{\ups}\|^2\bs{I}$, 
and MMSE reduces to a matched-filter, and estimation is as follows:
\begin{align} \bs{z}_k  \approx \bs{\Ups}^\t \bs{\Ups} \, \bs{c}_k + \bs{\Ups}^\t \bs{n}_k 
 = \|\bs{\ups}\|^2 \, \bs{c}_k + \bs{\nu}_k, \quad\qquad \bs{\nu}_k\sim\mathcal{N}(\bs{0},\sigma_{\_n}^2 \|\bs{\ups}\|^2\bs{I}). \label{eq:DLerror}
\end{align}
Dividing by $\|\bs{\ups}\|^2$ the previous expression yields:
\begin{equation} \bs{\hat{c}}_k \eqdef \frac{1}{\|\bs{\ups}\|^2} \bs{z}_k 
\approx \bs{c}_k + \frac{1}{\|\bs{\ups}\|} \bs{\nu}_k 
 = \bs{c}_k + \bs{\nu}'_k,
\end{equation}
where $\bs{\nu}'_k\sim\mathcal{N}(\bs{0},(\sigman^2/\|\bs{\ups}\|^2)\bs{I})$. Note that, as well-known (\eg \cite{biglieri:coding}), the estimation can be made as accurate as we desired by increasing $\|\bs{\ups}\|^2$. For antipodal sequences, say $\phi[i]\in\{-\msf{A}_t,\msf{A}_t\}$ with $\msf{A}_t>0$, the energy of the training sequence is $\msf{A}_t^2 N_t^\mrm{DL}$; therefore, $\|\bs{\ups}\|^2$ can be increased either by increasing power spent on training, that is, by increasing $\msf{A}_t$, or by increasing time spent for training, that is, by increasing $N_t^\mrm{DL}$, or both.


\bigskip

In the uplink training, the BS receives the superposition of the sequences of users each filtered by the corresponding channel, that is:
\begin{equation}\label{eq:ultraining} \bs{y}^\mrm{UL} = \sum_{k=1}^K \bs{\Ups}^\mrm{UL}_{\!k} \bs{c}_{k}+\bs{n} \equiv \bs{\Ups}^\mrm{UL}\bs{c}+\bs{n}, \end{equation}
having defined:
\begin{equation}\label{eq:blockult} \bs{\Ups}^\mrm{UL}\eqdef \sum_{k=1}^K \bs{e}_k^\t \otimes \bs{\Ups}_{\_k\.}^\mrm{UL}, \qquad \bs{c}\eqdef \sum_{k=1}^K \bs{e}_k \otimes \bs{c}_k. \end{equation}
As previously, the superscript $\mrm{UL}$ is dropped to unclutter notation. 

The goal of the BS is to linearly estimate $\bs{c}$ by observing $\bs{y}$, knowing $\bs{\Ups}$:
\begin{equation}\label{eq:estUL} \bs{z} = \bs{W}^\t \bs{y} \equiv \sum_{k=1}^K \bs{e}_k \otimes \bs{z}_k = \bs{W}^\t \bs{\Ups}\bs{c} + \bs{W}^\t \bs{n}, \end{equation}
where $\bs{z}$ is the $KL\ii\times 1$ vector of channel estimations, being $\bs{z}_k$ the $L\ii\times 1$ vector representing $\bs{c}_k$ estimate, and $\bs{W}^\t$ is the $KL\ii\times N_t^\mrm{UL}\ii$ matrix representing the estimator. All common linear estimators, that is ZF (Zero-Forcing), RZF (Regularized Zero-Forcing), MMSE (minimum mean square error) and MF (matched-filter), can be described by the following expression, parametrized by $\xi$ and $\zeta$: 
\begin{equation}\label{eq:generalult} \bs{W}^\t = (\xi\bs{\Ups}^\t\bs{\Ups}+\zeta\bs{I})^{-1} \bs{\Ups}^\t. \end{equation}
Indeed, MMSE is obtained with $(\xi,\zeta)=(1,\sigma_{\_n}^2)$; ZF with $(\xi,\zeta)=(1,0)$; MF with $(\xi,\zeta)=(0,1)$; RZF with $(\xi,\zeta)=(1,z)$.

In the simple case of ZF, the form assumed by eq.~\eqref{eq:estUL} is as follows:
\begin{equation}\label{eq:estULZF} \bs{z} = \bs{c}+(\bs{\Ups}^\t\bs{\Ups})^{-1}\bs{\Ups}^\t\bs{n} \equiv \bs{c}+\bs{\nu}, \quad \bs{\nu}\sim\mathcal{N}(\bs{0},\sigma_{\_n}^2(\bs{\Ups}^\t\bs{\Ups})^{-1}) \end{equation}
and, therefore, the $\ell$-th tap of the channel of generic user $k$ is:
\[ z_k[\ell] = c_k[\ell]+\nu_k[\ell]. \]
Here, $\nu_k[\ell]$ is a correlated Gaussian random variable with variance coinciding with the $((k-1)L\ii+\ell+1)$-th diagonal element of $\sigman^2(\bs{\Ups}^\t\bs{\Ups})^{-1}$. 

Assuming all UTs are transmitting the same power, \ie, $\|\bs{\ups}_k\|^2$ is the same for each $1\leq k\leq K$, the approximation $\bs{\Ups}_{\_j}^\t \bs{\Ups}_{\_i\.}=\|\bs{\ups}\|^2 \bs{I}\delta_{ji}$ allows to assume uncorrelated estimation errors, since $\bs{\Ups}^\t \bs{\Ups}=\|\bs{\ups}\|^2 \bs{I}$, and thus:
\begin{equation}\label{eq:zfapp} \bs{z}=\bs{c}+\bs{\nu},\quad \bs{\nu}\sim\mathcal{N}(\bs{0},(\sigman^2/\|\bs{\ups}\|^2)\bs{I}),\quad N_t^\mrm{UL}\gg L. \end{equation}




\subsection{Performance measures}
In both system structures, the statistic for inferring the transmitted symbol $b_k$ of user $k$ can be written in the following form:
\[ z_k = a_{kk} b_k + \sum_{\substack{j=1\\j\neq k}}^K a_{kj}b_j + \nu_k = a_{kk}b_k + I_k + \nu_k, \]
where $\nu_k$ is a r.v. representing noise, and $\{a_{kj}\colon j=1,\dotsc,K\}$ are r.vs. depending on multipath channels, random time-hopping codes, random delays, and estimation errors.

\smallskip

Two performance measures are considered. 

In the \textit{uncoded} regime, the probability of error as defined by:
\[ P_e = \frac{1}{2}\Prob{z_k<0\;\Big|\;b=\sqrt{\En}}+\frac{1}{2}\Prob{z_k>0\;\Big|\;b=-\sqrt{\En}}, \]
is considered.

In the \textit{coded} regime, mutual information with Gaussian inputs and a bank of matched-filters followed by independent decoders is considered; for the generic user $k$, this is given by:
\begin{equation}\label{eq:defIk} I(b_k; z_k) \quad \textup{nats/channel use},\end{equation}
where $I(b_k; z_k)$  is the mutual information between the transmitted symbol $b_k$ and the decision variable $z_k$. Since a channel use corresponds to $NT_\msf{c}=N\ii/W$ seconds, the sum-rate achieved by the set of $K$ users is:
\begin{equation}\label{eq:defRk} R\triangleq W\frac{\beta}{\ii} I(b; z) \quad \textup{nats/s},\end{equation}
having indicated with $I(b; z)$ the mutual information \eqref{eq:defIk} for a generic user. Finally, a spectral efficiency equal to
\begin{equation}\label{eq:defeffk} \mathcal{R}\triangleq \frac{\beta}{\ii} I(b; z) \quad \textup{(nats/s)/Hz}\end{equation}
is obtained.

\input{sec_prerr}

\section{Mutual information, Sum-Rate, and Spectral Efficiency}\label{sec:arates}

In this section, mutual information \eqref{eq:defIk} is derived for AR and TR. The other merit figures \eqref{eq:defRk} and \eqref{eq:defeffk} follows directly, although all the elements for a comparison are already included in \eqref{eq:defIk}. 

\subsection{Derivation of Mutual Information}

The decision variable for both the imperfect {TR} (c.f. eq.~\eqref{eq:zTRimperfect}) and {AR} (c.f. eq.~\eqref{eq:zARimperfect}) can be cast in the following form:
\begin{align}\label{eq:zTRimperfectSIMPL}
\hat{z}_{k} 
	 = \hat{a}_{kk} b_k + \sum_{\substack{j=1\\j\neq k}}^K \hat{a}_{kj} b_j + \hat{\nu}_k  = \hat{a}_{kk} b_k + \hat{S}_k + \hat{\nu}_k,
\end{align} 
where $\hat{\nu}_k=n_{q_k}\sim\mathcal{N}(0,\N/2)$. 

Let specify and give an interpretation of the terms $\hat{a}_{ki}$, $i=1,\dotsc,K$, for both TR and AR. 

\medskip
\textit{TR coupling coefficients.}

For TR, the term $\hat{a}_{kk}$ is given by:
\begin{align*}
\hat{a}_{kk}^\mrm{TR} 	= \bs{e}_{q_k}^\t \bs{C}_k \alpha_k (\bs{T}_{k}+\bs{\Xi}_k) \bs{x}_k 
				= \bs{c}_k^{\leftarrow\t} \frac{\bs{t}_k+\bs{\xi}_k}{\|\bs{t}_k+\bs{\xi}_k\|} = \bs{c}_k^{\t} \frac{\bs{c}_k+\bs{\xi}_k^\leftarrow}{\| \bs{c}_k+\bs{\xi}_k^\leftarrow \|},
\end{align*}
where $\bs{v}^{\leftarrow}\!\in\R^n$ denotes a vector with same components of vector $\bs{v}$ in reversed order, \ie, $[\bs{v}^{\leftarrow}]_i=\bs{v}_{n-i+1}$, $1 \leq i\leq n$. The term $\hat{a}_{kj}$, $j\neq k$, is:
\begin{align}
\hat{a}_{kj}^\mrm{TR}	 = \bs{e}_{q_k}^\t \alpha_j \bs{C}_j (\bs{T}_{j}+\bs{\Xi}_j) \bs{x}_j 
 					 = \bs{e}_{q_k}^\t \frac{1}{\|\bs{t}_j+\bs{\xi}_j\|} \bs{C}_j (\bs{t}'_j+\bs{\xi}'_j) 
					 = \bs{e}_{q_k}^\t \frac{1}{\|\bs{t}_j+\bs{\xi}_j\|} (\bs{\gamma}_j'+\bs{\zeta}_j'),
\end{align}
where $\bs{\gamma}_i'=[\bs{0}_{j_{\bs{x}_i}\_\times 1}^\t,\,\bs{\gamma}_i^\t,\,\bs{0}_{(N\ii-j_{\bs{x}_i})\times 1}^\t  ]^\t$ being $\bs{\gamma}_j$ the $(2(L+1)\ii-1)\times 1$ autocorrelation sequence of $(c_j[\ell])_{\ell=0}^{L\ii}$, and, similarly, $\bs{\zeta}_i'=[\bs{0}_{j_{\bs{x}_i}\_\times 1}^\t,\,\bs{\zeta}_i^\t,\,\bs{0}_{(N\ii-j_{\bs{x}_i})\times 1}^\t  ]^\t$ with $\bs{\zeta}_i$ a Gaussian random vector with non-identity correlation.

In order to provide an interpretation of the above expressions, it is useful to start with the case of no estimation error. In general, the decision variable at the output of the matched filter of user $k$ is given by the $q_k$-th sample of the sum of both intended and interference signals, plus noise. In the special case of no estimation errors, $a_{kk}=\|\bs{c}_k\|$ is the square root energy of channel $k$, \ie, the maximum tap of the effective channel, while $a_{kj}$ is either equal to zero if the effective channel of user $j$, that occupies $2(L+1)\ii-1$ out of $N\ii$ degrees of freedom in a symbol period, is not present at delay $q_k$, or to a random resolved path of the effective channel of user $j$, the randomness owing to random hopping and asynchronism. In presence of estimation errors, $\hat{a}_{kk}$ is smaller than, although in general in the neighbourhood of, the square root energy of channel $k$ due to the mismatch between $\bs{\hat{h}}_k$ and $\bs{h}_k$, and $\hat{a}_{kj}$ is either equal to zero if the \textit{perturbed} effective channel of user $j$, that occupies $2(L+1)\ii-1$ out of $N\ii$ degrees of freedom in a symbol period, is not present at delay $q_k$, or equal to a random path of the \textit{perturbed} effective channel of user $j$, the perturbation owing to the imperfect channel estimation of user $j$.

\medskip
\textit{AR coupling coefficients.}

For AR, the set of coupling coefficients $\{\hat{a}_{ki}\}_{i=1}^K$ are:
\begin{align}
\hat{a}_{kk}^\mrm{AR} 	& = \frac{\bs{\hat{h}}_k^\t}{\|\bs{\hat{h}}_k\|} \bs{h}_k = \|\bs{c}_k\|^2 + \bs{\xi}^\t_k \bs{c}_k, \label{eq:zAR1}\\ 
\hat{a}_{kj}^\mrm{AR} 	& = \frac{\bs{\hat{h}}_k^\t}{\|\bs{\hat{h}}_k\|} \bs{h}_j, \quad j\neq k \label{eq:zAR2}
\end{align}
where $\bs{\hat{h}}_k=\bs{h}_k+\bs{\chi}_k$, $\bs{h}_k=\bs{C}_k\bs{x}_k$, and $\bs{\chi}_k=\bs{\Xi}_k\bs{x}_k$. We can think of $\bs{\hat{h}}_k/\|\bs{\hat{h}}_k\|$ as the perturbed direction along which the received signal is projected in order to decode user $k$; $\hat{a}_{kk}^\mrm{AR}$ represents, therefore, the ``mismatch'' between the perturbed and unperturbed channels of user $k$; $\hat{a}_{kj}^\mrm{AR}$ represents the coupling between user $k$, that is perturbed, and another user $j$. As in the TR case, a channel impulse response occupies a fraction, that is approximately equal to $(L+1)/N$, of the available degrees of freedom in a symbol period; Opposite of the TR case, where the perturbation affects user $j$ in $\hat{a}_{kj}^\mrm{TR}$, user $k$ is perturbed in the AR case (through $\bs{\hat{h}}_k$), user $j$ appearing with the true channel impulse response $\bs{{h}}_j$.

\begin{figure}[t]
\centering
\setlength\ploth{0.375\columnwidth} 
\setlength\plotw{0.425\columnwidth}
\subfigure[\emph{$\sigma_\xi^2=0$}]{\label{fig:01}  \includegraphics[scale=0.9]{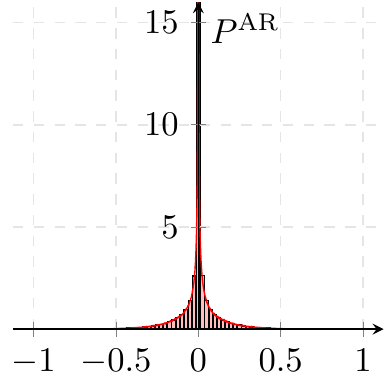} }
\subfigure[\emph{$\sigma_\xi^2=0$}]{ \label{fig:02}\includegraphics[scale=0.9]{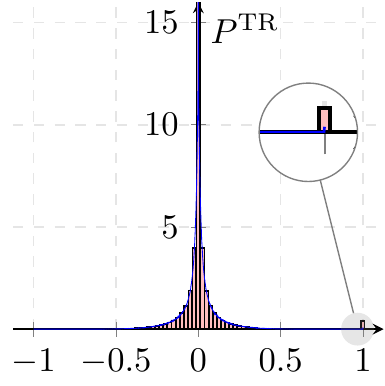} } 
\subfigure[\emph{$\sigma_\xi^2=0.01$}]{\label{fig:03}  \includegraphics[scale=0.9]{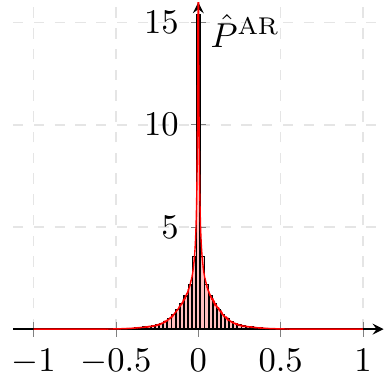} }
\subfigure[\emph{$\sigma_\xi^2=0.01$}]{ \label{fig:04}\includegraphics[scale=0.9]{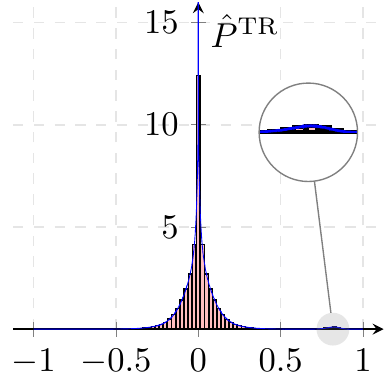} } \\
\subfigure[\emph{$\sigma_\xi^2=0.01$}]{ \label{fig:05}\includegraphics[scale=0.9]{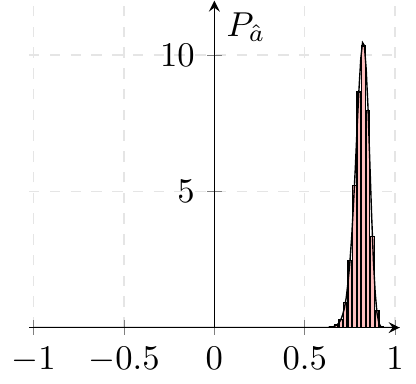} }
\caption{PDFs $\hat{P}$ and $P_{\hat{a}}$ of the variables $\hat{a}_{kj}$, $j\neq k$, and $\hat{a}_{kk}$, respectively. Distributions do not depend on the particular user $k$. Figs. \textit{(a)} and \textit{(b)} correspond to a variable $\hat{a}_{kj}$, $j\neq k$, in systems without estimation errors; Figs. \textit{(c)} and \textit{(d)} correspond to a variable $\hat{a}_{kj}$, $j\neq k$, in systems with estimation errors with variance $\sigma_\xi^2=0.01$; Fig. \textit{(e)} corresponds to a variable $\hat{a}_{kk}$ in systems with a same error variance.}
\label{fig:0104}
\end{figure}

\medskip
\textit{Derivation.}

Being each term in the r.h.s. of eq.~\eqref{eq:zTRimperfectSIMPL} independent from the other terms, mutual information $I(z_k;b_k)$ can be derived once the distributions of $\hat{a}_{kk}$ and $\hat{S}_k$ are known. The former depends on both the random channel impulse response and estimation errors of user $k$, and the latter on the random channel impulse responses and estimation errors, and the random delays with respect to user $k$. Hence, the final form assumed by $I(z_k;b_k)$ strongly depends on the channel model; however, in the following, the effect of the time-hopping and random asynchronism will be enucleated, without enter in the computation of a mutual information when a particular channel model is adopted; this last task is addressed by simulations, where the IEEE 802.15.3a model \cite{foerster:chmodel}, that is valid for bandwidths up to several gigahertz, is selected. 


As for TR, since the effective channel of user $j$ occupies a fraction $\pp=(2(L+1)\ii-1)/(N\ii)$, and user $k$, due to the assumptions on independence and uniformity of hopping codes and asynchronism, selects uniformly at random one of the $N\ii$ samples available per symbol period, then ${a}_{kj}$, $j\neq k$, is equal to zero with probability $1-\pp$, and is distributed as the generic path of the effective channel $\bs{{h}}_j$ with probability $\pp$, that is:
\begin{equation}\label{eq:TRsinoerror} P_{{a}_{kj}^\mrm{TR}}= (1-\pp)\delta_0+\pp {P}^\mrm{TR}, \end{equation}
where ${P}^\mrm{TR}$ indicates the distribution of the generic path of the effective channel of user $j$ (that is independent of $j$). In presence of estimation errors, the above argument holds, that is, $\hat{a}_{kj}$, $j\neq k$, is equal to zero with probability $1-\pp$, and is distributed as the generic path of the \textit{perturbed} effective channel $\bs{\hat{h}}_j$ with probability $\pp$:
\begin{equation}\label{eq:TRsi} P_{\hat{a}_{kj}^\mrm{TR}}= (1-\pp)\delta_0+\pp \hat{P}^\mrm{TR}, \end{equation}
where $\hat{P}^\mrm{TR}$ indicates the distribution of the generic path of the \textit{perturbed} effective channel of user $j$ (that is independent of $j$). 

As for AR, let start by finding the distribution of ${a}_{kj}$, $j\neq k$, \ie, the coupling coefficient between two users in absence of estimation error. Each channel spans a subspace of dimension $(L+1)\ii$ in a space with $(N+L+1)\ii-1$ dimensions\footnote{The number of degrees of freedom in a symbol period is $N\ii$; in the large system limit, as $N\gg L$, the difference between $(N+L+1)\ii-1$ and $N\ii$ due to the convolution is negligible.}; in other words, just $(L+1)\ii$ entries of $\bs{h}_i$ are nonzero. From the hypotheses of independence and uniformity of delays due to asynchronism between users and time-hopping codes, there exists a probability $f$ such that the inner product $\bs{h}_k^\t\bs{h}_j$ is nonzero, and the remaining probability $1-f$ that the inner product is zero. We may think of the ``nonzero event'' as the partial overlapping between two channels. As $N\gg L$, it results $f\approx (2(L+1)\ii-1)/(N\ii)$, where the assumption $N\gg L$ allows to neglect border effects. Indicating with $P^\mrm{AR}$ the distribution of ${a}_{kj}^\mrm{AR}$ conditioned on the nonzero event, one has:
\begin{equation}\label{eq:ARsinoerror} P_{{a}_{kj}^\mrm{AR}} = (1-f) \delta_0 + f P^\mrm{AR}. \end{equation}
In presence of estimation errors, the above discussion remains valid, since an error $\bs{\chi}_k$ changes, in general, the direction of vector $\bs{\hat{h}}_k$ with respect to $\bs{h}_k$, \ie, $\bs{\hat{h}}_k$ and $\bs{{h}}_k$ are, in general, not collinear, but it does not change the subspace spanned by the two channels, \ie, the subspace spanned by the true channel is equal to the subspace spanned by the perturbed channel. Indicating with $\hat{P}^\mrm{AR}$ the distribution of $\hat{a}_{kj}^\mrm{AR}$ conditioned on the nonzero event, one has:
\begin{equation}\label{eq:ARsi} P_{\hat{a}_{kj}^\mrm{AR}} = (1-f) \delta_0 + f \hat{P}^\mrm{AR}. \end{equation}
$\hat{P}^\mrm{AR}$ reduces to ${P}^\mrm{AR}$ when the estimation error is nil.

Figure~\ref{fig:0104} shows the distributions $P^\mrm{TR}$, $\hat{P}^\mrm{TR}$, $P^\mrm{AR}$ and $\hat{P}^\mrm{AR}$ in eqs. \eqref{eq:TRsinoerror}, \eqref{eq:TRsi}, \eqref{eq:ARsinoerror} and \eqref{eq:ARsi}, respectively, and the distribution $P_{\hat{a}}$ of the term $\hat{a}_{kk}$, assuming the Channel Model 1 (CM1) specified in the IEEE 802.15.3a standard. All simulations refer to a system with fixed chip duration $T_\msf{c}=1$ ns and bandwidth $\msf{W}=1/T_\msf{c}$. Power control is assumed; in particular, $\|\bs{c}_k\|^2=1$ for all users, $k=1,\dotsc,K$. The delay spread of each channel impulse response is fixed at a value $T_d=50$ ns that includes most of the energy of typical CM1 channels. For a given bandwidth $\msf{W}$, the length of the channel expressed in number of samples per channel is $T_d\msf{W}$, \ie, $\bs{c}_k$ is a $(\lfloor T_d\msf{W}\rfloor+1)\times 1$ vector. 

\begin{figure*}
\centering
\setlength\ploth{0.7\columnwidth} 
\setlength\plotw{0.85\columnwidth}
\subfigure[\emph{$\sigma_\xi^2=0$, $\SNR=0$ dB.}]{\label{fig:06}  \includegraphics[scale=0.9]{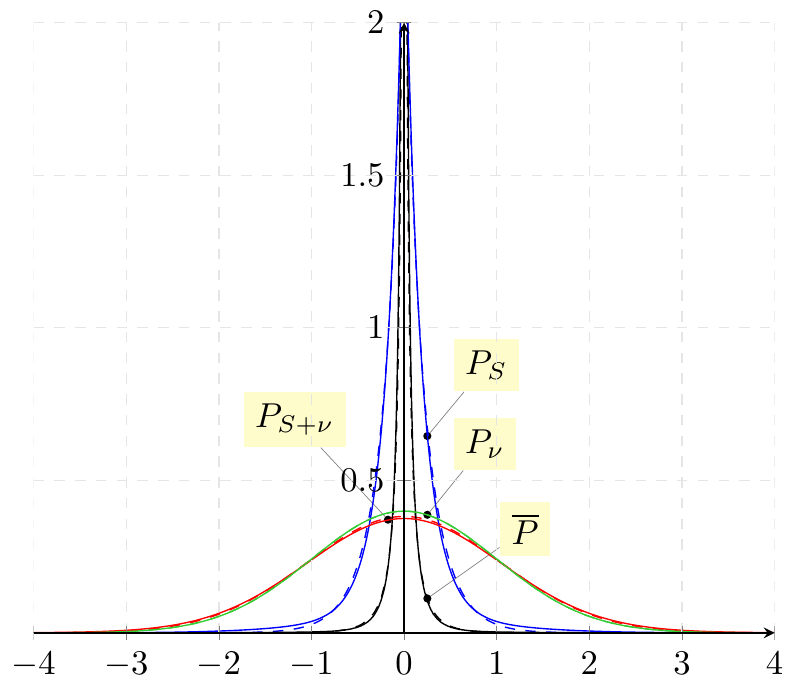} } \hfill
\subfigure[\emph{$\sigma_\xi^2=0.02$, $\SNR=0$ dB.}]{ \label{fig:07}\includegraphics[scale=0.9]{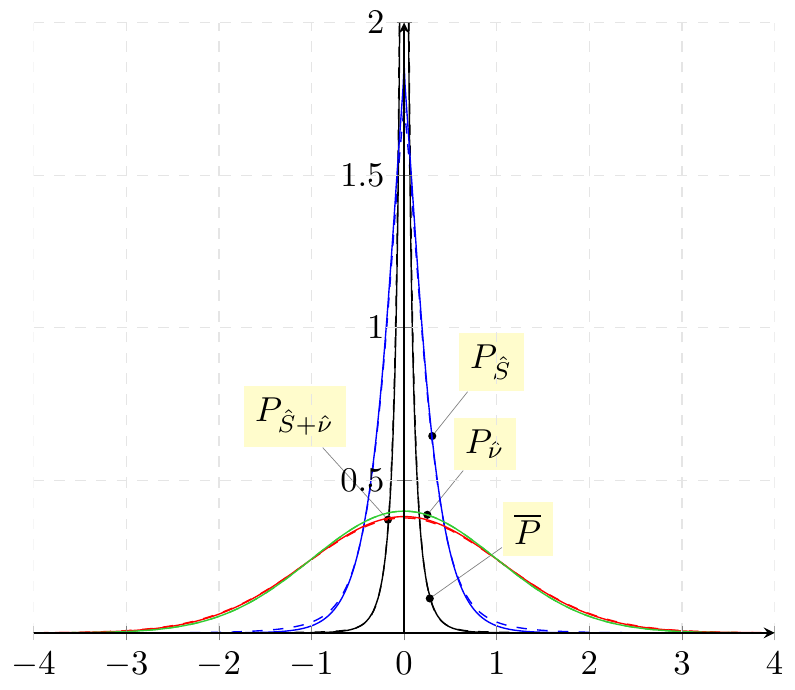} } \\
\subfigure[\emph{$\sigma_\xi^2=0$, $\SNR=20$ dB.}]{\label{fig:08}  \includegraphics[scale=0.9]{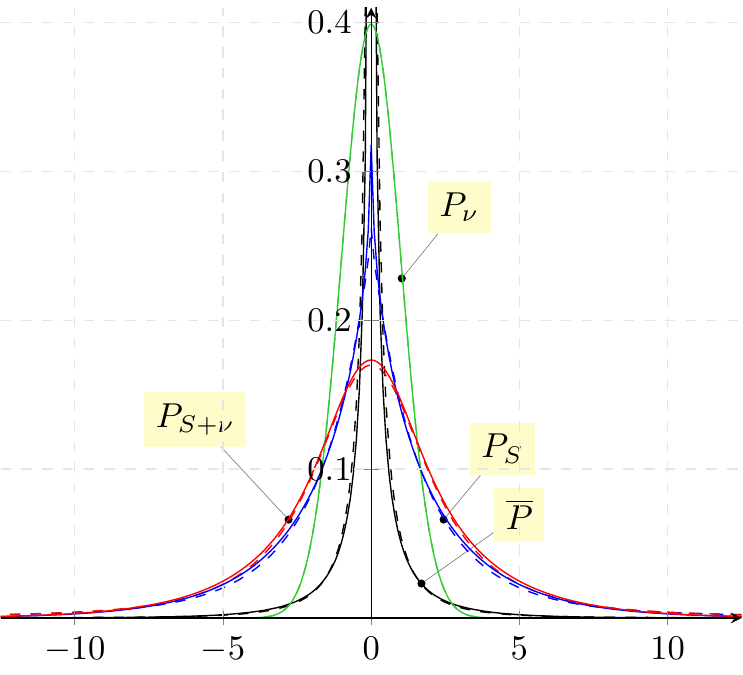} } \hfill
\subfigure[\emph{$\sigma_\xi^2=0.02$, $\SNR=20$ dB.}]{ \label{fig:09}\includegraphics[scale=0.9]{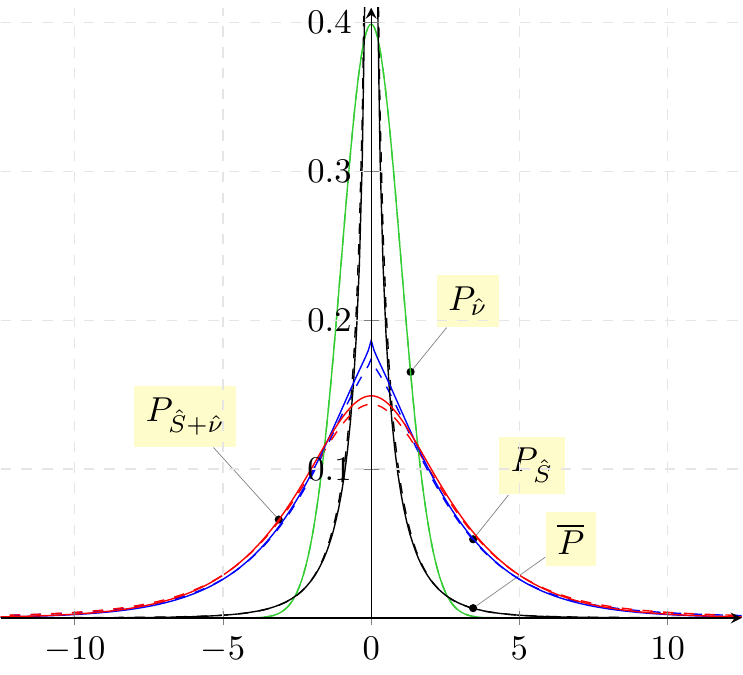} }
\caption{PDF $\ol{P}$ of interfering terms $\hat{a}_{kj}b_j$, $j\neq k$, given $\hat{a}_{kj}\neq 0$, $P_{\hat{\nu}}$ of the noise variable $\hat{\nu}_{k}$, $P_{\hat{S}}$ of interference $\hat{S}_k$, and $P_{\hat{S}+\hat{\nu}}$ of interference-plus-noise $\hat{S}_k+\hat{\nu}_k$. Error variance $\sigma_\xi^2$ and $\SNR=\En/\sigman^2$ are specified below each subfigure. The noise variance $\sigman^2$ is normalized to $1$. The load $\beta$ of the network is fixed and equal to $\beta=0.1$. Solid vs. dashed curves refer to TR vs. AR, respectively.}
\label{fig:0607}
\end{figure*}

Figures~\ref{fig:0104} \textit{(a)} vs. \textit{(b)} show the distributions of the coupling coefficient $a_{kj}$, $j\neq k$, in case of no estimation errors, for AR and TR, respectively. As may be expected, the variance of the latter is larger than the variance of the former, as follows from the property of time reversal to increase the total energy of the effective channel; In the specific case, the $\mrm{Var}[{a_{kj}^\mrm{AR}}]\approx 0.0099$ while $\mrm{Var}[{a_{kj}^\mrm{TR}}]\approx 0.0173$. In Fig.~\ref{fig:0104} \textit{(b)} it is highlighted the presence of a strong interference ($a_{kj}^\mrm{TR}=1$) that is not present in the AR case; in the TR case, there is, indeed, among the $2(L+1)\ii-1$ paths of the effective channel, one path with amplitude equal to the square-root energy of the channel, $\|\bs{c}_k\|=1$, that is, therefore, selected with probability $1/(2(L+1)\ii-1)$. $P^\mrm{TR}$ is also more leptokurtic than $P^\mrm{AR}$, showing a kurtosis approximately equal to $34$ vs. $12$ reached by the AR variable. Figures~\ref{fig:0104} \textit{(c)} vs. \textit{(d)} show the distributions of the coupling coefficient $\hat{a}_{kj}$, $j\neq k$, in case of estimation error with per sample variance $\sigma_\xi^2=0.01$, for AR and TR, respectively. Both variance and kurtosis of TR are still larger than those of AR; in particular, it results $\mrm{Var}[{a_{kj}^\mrm{AR}}]\approx 0.0099$ vs. $\mrm{Var}[{a_{kj}^\mrm{TR}}]\approx 0.0149$, and $\mrm{kurt}[{a_{kj}^\mrm{AR}}]\approx 6.9$ vs. $\mrm{kurt}[{a_{kj}^\mrm{TR}}]\approx 21.7$. Figure~\ref{fig:0104} \textit{(e)} shows the distribution of the term $\hat{a}_{kk}$, that is the same for both TR and AR, and is represented for $\sigma_\xi^2=0.01$. Note that all variables $\hat{a}_{ki}$ can be regarded as normalized cross-correlations, hence $-1 \leq \hat{a}_{ki}\leq 1$.

\begin{figure*}
\centering
\setlength\ploth{0.65\columnwidth} 
\setlength\plotw{0.9\columnwidth}
\subfigure[\emph{$\ii=1$, $\beta=0.1$, $\sigma_\xi^2=0$.}]{\label{fig:MI_10} \includegraphics[scale=0.85]{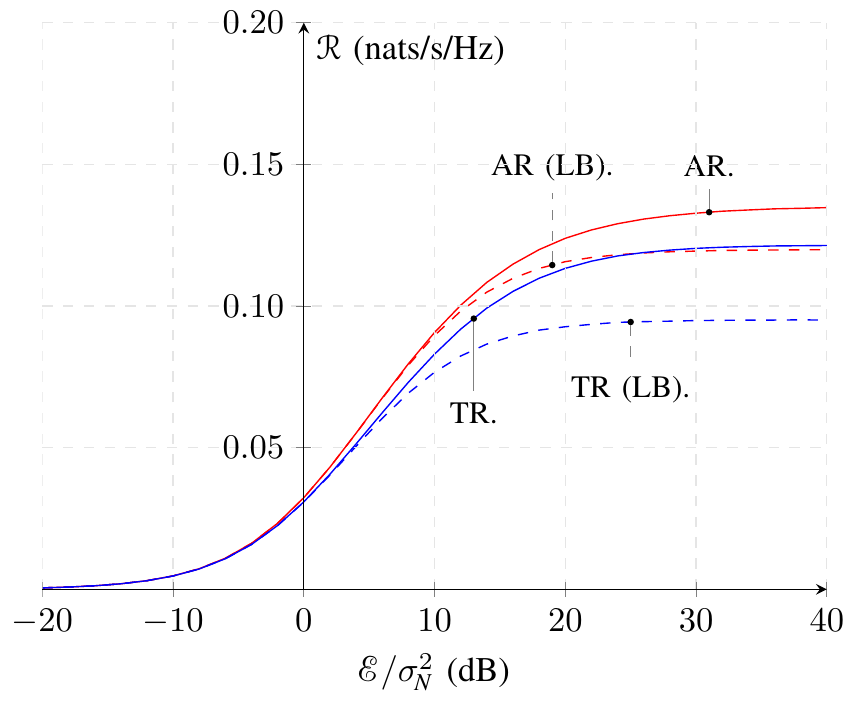} } \hfill 
\subfigure[\emph{$\ii=1$, $\beta=0.2$, $\sigma_\xi^2=0$.}]{\label{fig:MI_11} \includegraphics[scale=0.85]{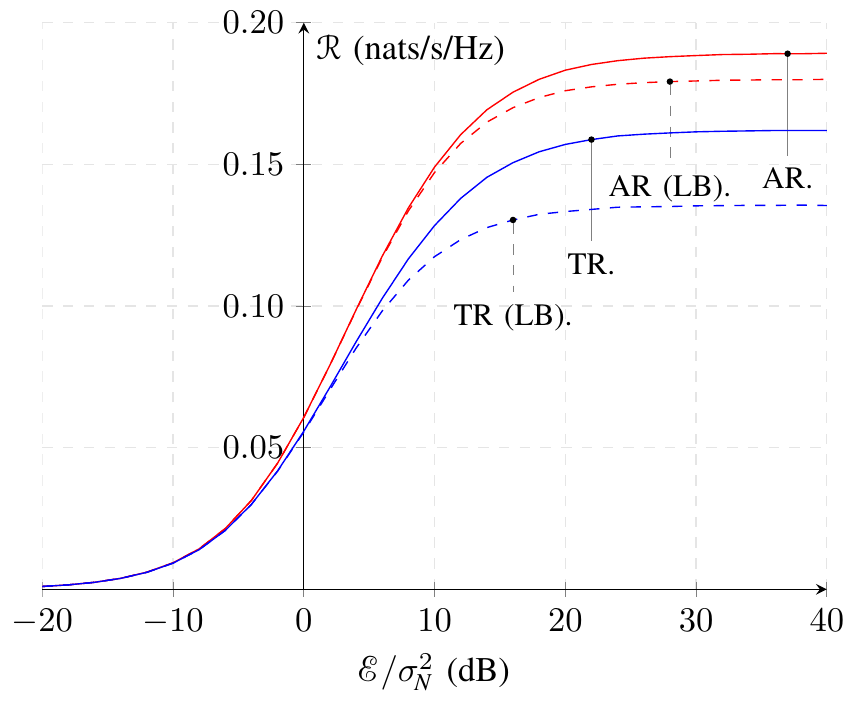} } \\
\subfigure[\emph{$\ii=1$, $\beta=0.1$, $\sigma_\xi^2=0.02$.}]{\label{fig:MI_12}  \includegraphics[scale=0.85]{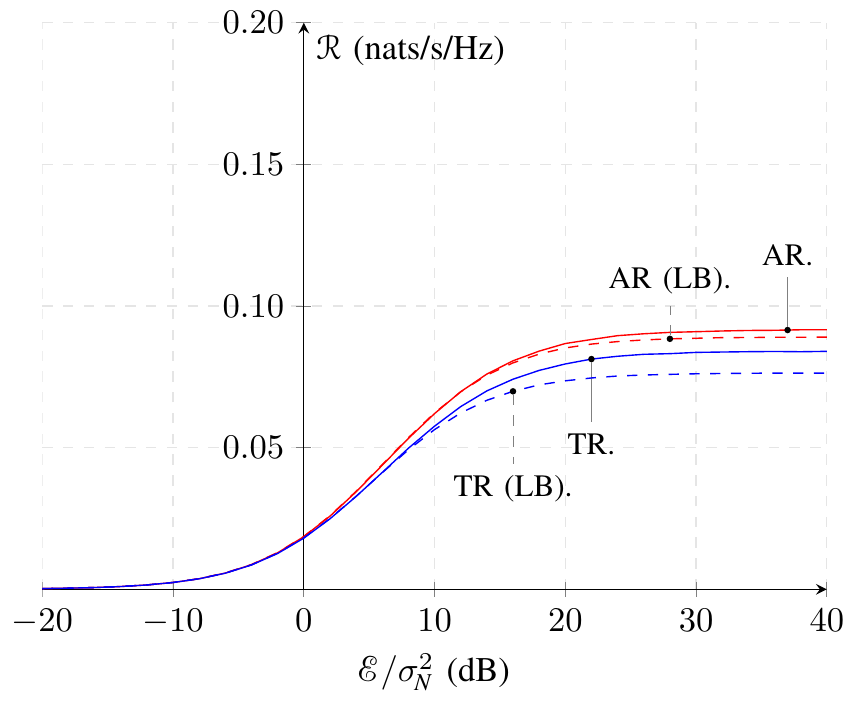} } \hfill
\subfigure[\emph{$\ii=1$, $\beta=0.2$, $\sigma_\xi^2=0.02$.}]{ \label{fig:MI_13}\includegraphics[scale=0.85]{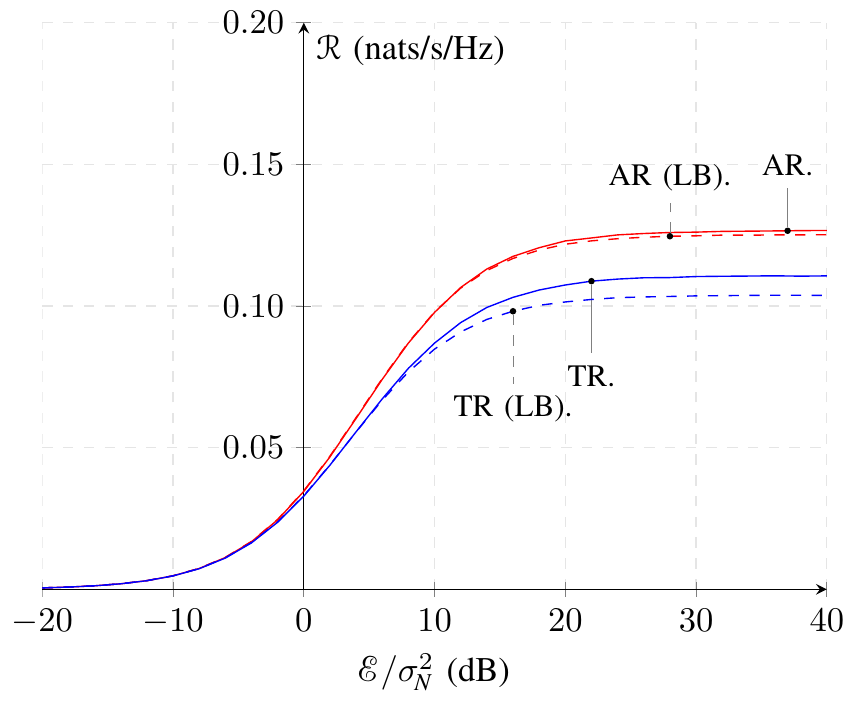} } \\
\subfigure[\emph{$\ii=2$, $\beta=0.1$, $\sigma_\xi^2=0$.}]{\label{fig:MI_14}  \includegraphics[scale=0.85]{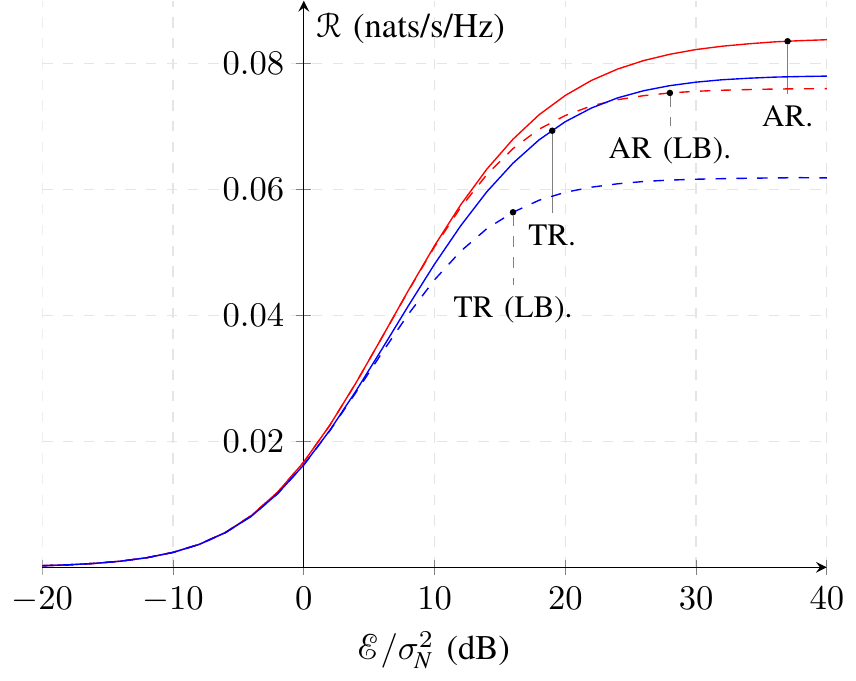} } \hfill
\subfigure[\emph{$\ii=2$, $\beta=0.1$, $\sigma_\xi^2=0.02$.}]{ \label{fig:MI_15}\includegraphics[scale=0.85]{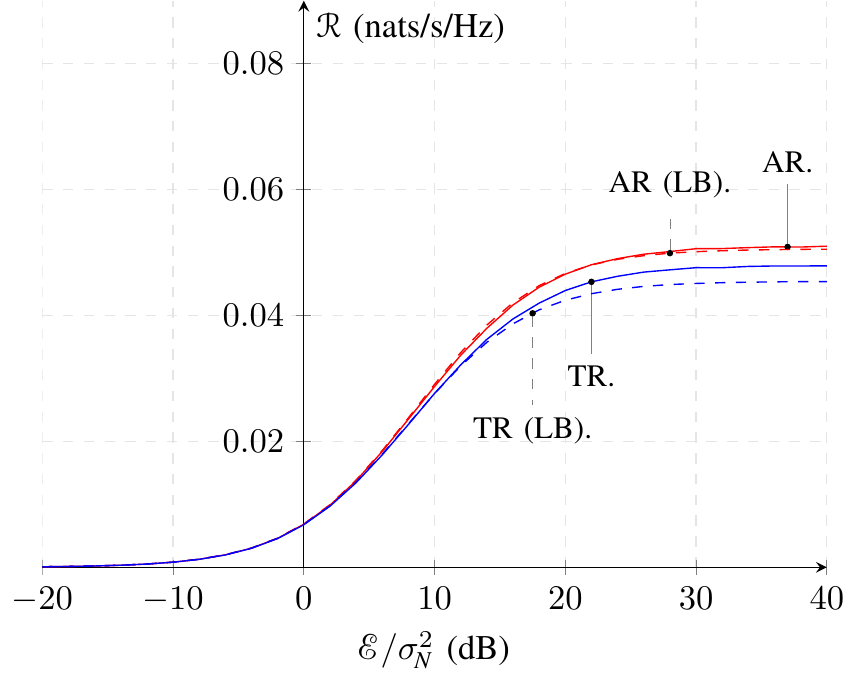} }
\caption{Spectral efficiency $\mathcal{R}$ (solid curve) and lower bound assuming Gaussian interference (dashed curve) vs. $\SNR=\En/\sigman^2$ (dB) for different values of load $\beta$ and error variance $\sigma_\xi^2$. Impulsiveness index is fixed to $\ii=1$ in Figs. from \textit{(a)} to \textit{(d)}, and to $\ii=2$ in Figs. \textit{(e)} and \textit{(f)}.}
\label{fig:MI}
\end{figure*}

\smallskip
In terms of c.fs., eqs.~\eqref{eq:ARsi} and \eqref{eq:TRsi} becomes:
\[ \varphi_{\hat{a}_{kj}}(u)=\E{e^{ju\alpha}} = 1-\pp( 1-\hat{\varphi}(u) ),\qquad \hat{\varphi}(u) = \int_\R d\alpha \, e^{j\alpha u}\hat{P}(\alpha), \]
being $\hat{P}$ equal to either $\hat{P}^\mrm{AR}$ or $\hat{P}^\mrm{TR}$ in eqs.~\eqref{eq:ARsi} and \eqref{eq:TRsi}, respectively. In general, given two independent r.vs. $X$ and $Y$ and their product $Z=XY$, it results $\varphi_Z(u)=\E{\varphi_Y(uX)}=\E{\varphi_X(uY)}$; therefore, the r.v. $\hat{a}_{kj}b_j$ has c.f.:
\[ \varphi_{\hat{a}_{kj}b_j}(u)= \E{ \varphi_{\hat{a}_{kj}}(b_j u) } = 1-\pp( 1-\E{\hat{\varphi}(b_j u)} )=1-\pp(1-\ol{\varphi}(u)),  \]
where the expectation is with respect $b_j\sim\mathcal{N}(0,\En)$, and $\ol{\varphi}(u)$ is independent of $j$. Since $\{\hat{a}_{kj}b_j\}$ with $j\in\{1,\dotsc,K\}\setminus k$ are independent, then $\hat{S}_k$ has c.f.:
\[ \varphi_{\hat{S}_{k}}(u) = \left\{1-\frac{2(L+1)\ii-1}{N\ii}(1-\ol{\varphi}(u))\right\}^{\!K-1},\]
that, in the large system limit, where $K\to\infty$, $N\to\infty$, $K/N\to\beta$, converges to:
\[
\varphi_{\hat{S}_{k}}(u) 	 = e^{-\beta_\mrm{eff}[1-\ol{\varphi}(u)]}
				 = \sum_{r\geq 0} \frac{\beta_\mrm{eff}^r}{r!} e^{-\beta_\mrm{eff}} [\ol{\varphi}(u)]^r 
,\]
where $\beta_\mrm{eff}=\beta(2(L+1)\ii-1)/\ii$ is the effective load; Without multipath ($L=1$) and one pulse per chip ($\ii=1$), $\beta_\mrm{eff}$ reduces to the usual load $\beta$ as given by $K/N$. The interference-plus-noise variable has thus c.f. given by:
\begin{align*} \varphi_{\hat{S}_k+\hat{\nu}_k}(u) & = \varphi_{\hat{S}_k}(u)\varphi_{\hat{\nu}_k}(u)  = e^{-\beta_\mrm{eff}[1-\ol{\varphi}(u)]-\frac{\sigman^2}{2}u^2}. \end{align*}

Figure~\ref{fig:0607} shows distributions $\ol{P}$, $P_{\hat{S}}$, $P_{\hat{\nu}}$, and $P_{\hat{S}+\hat{\nu}}$, that correspond to c.fs. $\ol{\varphi}$, $\varphi_{\hat{S}}$, $\varphi_{\hat{\nu}}$, and $\varphi_{\hat{S}+\hat{\nu}}$ defined above, for different values of $\sigma_\xi^2$ and $\SNR=\En/\sigman^2$, and fixed value of load $\beta=0.1$. Figures~\ref{fig:0607} \textit{(a)} and \textit{(b)} show a noise-limited scenario, where $\SNR=0$ dB, without and with estimation errors, respectively: simulations show that the interference-plus-noise variable is not significantly affected by estimation errors (c.f. $P_{{S}+{\nu}}$ and $P_{\hat{S}+\hat{\nu}}$ curves), although $P_S$, as well as $\ol{P}$, becomes less leptokurtic in presence of estimation errors. Figures~\ref{fig:0607} \textit{(a)} and \textit{(b)} show an interference-limited scenario, where $\SNR=20$ dB: in this case $P_S$ is far from Gaussian, and so is the interference-plus-noise PDF $P_{S+\nu}$; the effect of the estimation error is to decrease the kurtosis of $P_S$, and so that of $P_{S+\nu}$.

Knowing the distribution of $z_k$ conditioned on $b_k$, or equivalently its c.f., mutual information $I(z_k;b_k) = h(z_k)-h(z_k|b_k)$ follows directly. Hence, the c.f. of $z_k$ given $b_k$ is:
\[ \varphi_{\hat{z}_k|b_k}(u) = \varphi_{\hat{S}_k+\hat{\nu}_k}(u)\varphi_{\hat{a}_{kk}}(b_k u); \]
hence, the c.f. of $\hat{z}_k$ is:
\[ \varphi_{\hat{z}_k}(u) = \E{ \varphi_{\hat{z}_k|b_k}(u) } = \varphi_{\hat{S}_k+\hat{\nu}_k}(u) \E{ \varphi_{\hat{a}_{kk}}(b_k u) },\]
where the expectation is over $b_k\sim\mathcal{N}(0,\En)$. Explicitly, one has:
\[ h(\hat{z}_k) = \int_\R -P_{\hat{z}_k}(z) \ln P_{\hat{z}_k}(z) dz, \quad P_{\hat{z}_k}(z) = \frac{1}{2\pi}\int_\R du\, e^{-j u z} \varphi_{\hat{z}_k}(u), \]
and:
\[ h(\hat{z}_k|b_k) = \int_\R \Phi_{0,\En}(b) h(\hat{a}_{kk}b+\hat{S}_k+\hat{\nu}_k)  = \int_\R \Phi_{0,\En}(b) \int_\R -P_{\hat{z}_k|b_k=b}(z) \ln P_{\hat{z}_k|b_k=b}(z) dz, \]
where $\Phi_{0,\En}$ denotes a Gaussian distribution with zero mean and variance $\En$, and:
\[ P_{\hat{z}_k|b_k=b}(z) = \frac{1}{2\pi}\int_\R du\, e^{-j u z} \varphi_{\hat{z}_k|b_k=b}(u). \]

%

The above derivation allows to find $I(\hat{z}_k;\hat{b}_k)=h(\hat{z}_k)-h(\hat{z}_k|b_k)$ as a function of the distribution $\hat{P}$ of $\hat{a}_{kj}$, $j\neq k$, and the distribution $P_{\hat{a}_{kk}}$ accounting for the loss of correlation incurred by the user to be decoded because of the estimation error. Both $\hat{P}$ and ${P}_{\hat{a}_{kk}}$ accounts for the channel model and the estimation error, in particular its variance. 

As baseline comparison, we also provide the following lower bound $\ul{I}(\hat{z}_k;b_k)$ for ${I}(\hat{z}_k;b_k)$, that is achieved when the interference is Gaussian:
\[	I(\hat{z}_k;b_k) 	 \geq \ul{I}(\hat{z}_k;b_k) 
				= {\Expectation}\,\bigg[\.{ \frac{1}{2}\ln\left(1+\frac{\En\.\hat{a}_{kk}^2}{\mrm{Var}[\,{\hat{S}_k}\,]+\Var{\hat{\nu}_k}}\right)}\.\bigg],
\]
where the expectation is over $\hat{a}_{kk}$. The corresponding spectral efficiency lower bound is $\ul{\mathcal{R}}=(\beta/\ii)\ul{I}(\hat{z}_k;b_k)$ (c.f. eq.~\eqref{eq:defeffk}). 

Results are shown on Fig.~\ref{fig:MI}, where $\mathcal{R}$ (solid curve) and $\ul{\mathcal{R}}$ (dashed curve) are presented as a function of $\En/\sigman^2$, for different values of $\beta$ and $\sigma_\xi^2$. The receiver structure shows a mutual information floor at high SNR. By comparing Figs.~\ref{fig:MI} \textit{(a)} and \textit{(b)}, one observes that $\mathcal{R}$ increases sublinearly as $\beta$ increases, while by comparing Figs.~\ref{fig:MI} \textit{(a)} and \textit{(c)}, or Figs.~\ref{fig:MI} \textit{(b)} and \textit{(d)}, a reduction in spectral efficiency due to the presence of an estimation error is observed. $\mathcal{R}$ scales with $\ii$ as shown on Figs.~\ref{fig:MI} \textit{(e)} and \textit{(f)}. 

In each of these simulations, AR outperforms TR. However, note that the gap $\mathcal{R}-\ul{\mathcal{R}}$ may be viewed as a measure of the nonGaussianity of $\hat{z}_k$ and $\hat{z}_k|b_k$, and is indeed higher in the TR case with respect to the AR case, because of the different distribution of the interference term, that is more leptokurtic in the TR case. There could be, therefore, a room for TR to outperform AR. 

\section{Conclusions and Future Investigations}\label{sec:conc}

In this paper, the problem of characterizing system performance for single antenna systems using time reversal in the case of imperfect channel estimation was addressed. The analyzed setting included one BS and several UTs, and the uplink communication channel was considered in the investigation. Each UT adopted impulse-radio ultra-wideband communication with prefiltering, and the receiving BS adopted a 1Rake; degrees of impulsiveness were reflected by an impulsiveness index that ranges from $\ii=1$ to $\ii\to\infty$ for ideal impulsiveness. In order to evaluate time reversal behavior, this communication scheme was compared against a reference configuration with no prefiltering and AR at the receiver. The effect of imperfect channel estimation on both transceiver configurations was analyzed. Channel estimation error was modeled as an additive Gaussian noise based on the output of a training phase that was used to tune transmitter and receiver structures. The comparison was performed for both the single user channel and the multiuser channel with power control. Modeling of the channels was obtained based on the 802.15.3a CM1 model. The two communication schemes, TR and AR, were compared based on two different performance parameters: probability of error and mutual information as a function of signal-to-noise ratio.

Results highlighted that, for the single user channel, probability of error for TR and AR coincided, while for the multiuser channel, AR outperformed TR when imperfect CSI was the main cause of error, and the two schemes had similar performance when the load, as measured by the ratio between the number of terminals $K$ and the number of chips $N$ in a symbol period, $\beta=K/N$, was the main cause of error, irrespectively of the degree of impulsiveness.

In terms of spectral efficiency, we provided lower and upper bound expressions, and analyzed the two structures with different impulsiveness index $\ii$ and load $\beta$. Results expressed by spectral efficiency $\mathcal{R}$ (nats/s/Hz) as a function of signal-to-noise ratio indicated that, for low-SNR, $\mathcal{R}$ was similar for the two systems, while for higher SNR values, AR outperforms TR. However, remind that, in practical scenarios, it would be simpler for a TR system to acquire a better estimation of the channel with respect to an AR system, since the estimation error variance depends on the energy of the training sequence only, and in the TR case the training sequence is transmitted by a basestation rather than a device, and may require weaker energy consumption constraints. Furthermore, in the presence of estimation errors, a reduction in $\mathcal{R}$ was observed, due to both a mismatch with the user to decode, and a reduced kurtosis of the interference term. 


Future investigations should explore extensions of the present model by removing the hypothesis of a single user detector, \ie, increasing the complexity in the receiver to multiuser detection, and by considering different path losses characterizing the different channels of the different links, \ie, removing the unit gain assumption that makes all transmitted power equal.

\section*{Acknowledgment}
This work was partly supported by European Union and European Science Foundation under projects COST Action IC0902 and FP7 Network of Excellence ACROPOLIS. G.C.F. acknowledges Sapienza University of Rome for the PhD Institutional Fellowship, as well as Research Grant, Anno 2013 - prot. C26A13ZYM2, and the Project ICNET.

\normalsize
\ifCLASSOPTIONcaptionsoff
  \newpage
\fi

\end{document}

%% file: myshort.tex
\usepackage{xcolor,color,graphicx}
\usepackage[bookmarks=false,colorlinks,
			pdfauthor={Guido Carlo Ferrante},%
			pdftitle={TWC}]{hyperref}%
						\hypersetup{colorlinks,%
						citecolor= blue,%
						filecolor= blue,%
						linkcolor= black,%
						urlcolor= blue}
\usepackage{upgreek}

\newtheoremstyle{classicthm}
  {4pt}
  {4pt}
  {\itshape}
  {}
  {\itshape}
  {.}
  { }
  {}

\theoremstyle{classicthm}

\theoremstyle{remark}

\usepackage{braket}
\usepackage{booktabs}
\usepackage{caption}
\usepackage[mathcal]{euscript}
\usepackage[scr=rsfso,bb=boondox,frak=euler,scaled=1.035]{mathalfa}
\usepackage{rotating}
\usepackage{microtype,ragged2e}

\usepackage{tikz}
\usetikzlibrary{arrows,decorations,decorations.pathmorphing,decorations.pathreplacing,%
backgrounds,positioning,fit,matrix,snakes,calc,plotmarks,patterns,spy,shadows}
\usepackage{pgf,pgfbaseimage,pgffor}
\usepackage{pgfplots}
\usepackage{tkz-fct}
\usepackage{enumerate}
\newcommand{\frenchit}[1]{\text{\usefont{T1}{frc}{m}{it}#1}}
\def\ii{\frenchit{i}}

\def\pp{f}

\usepackage{multirow}

\usepackage{bm}
\newcommand{\bs}[1]{\bm{#1}}
\newcommand{\numberset}[1]{\mathbb{#1}} 
\newcommand{\gT}[1]{\mrm{g}_{#1}}

\def\ups{\upsilon}
\def\Ups{\Upsilon}

\def\En{\mathscr{E}}
\def\sigman{\sigma_{\_\textit{N}}}
\newcommand{\UT}{\textup{UT}}

\newcommand{\R}{\numberset{R}}
\newcommand{\Z}{\numberset{Z}}
\newcommand{\N}{\hspace{-1pt}\mathscr{N}_{0}}

\DeclareMathOperator{\Expectation}{\mathbb{E}\!}
\newcommand{\E}[1]{\Expectation\left[\,#1\,\right]}
\newcommand{\Var}[1]{\mrm{Var}\!\left[\,#1\,\right]}

\newcommand{\Prob}[1]{\mathbb{P}\!\left(\,#1\,\right)}

\newcommand{\ol}[1]{\overline{#1}}
\newcommand{\ul}[1]{\underline{#1}}

\def\t{\msf{T}}
\newcommand{\SNR}{\mathsf{SNR}}
\newcommand{\SINR}{\mathsf{SINR}}

\def\.{\hspace{1pt}}
\def\_{\hspace{-1pt}}
\def\+{\hspace{-1pt}+\hspace{-1pt}}
\def\-{\hspace{-1pt}-\hspace{-1pt}}
\def\={\hspace{-1pt}=\hspace{-1pt}}
\def\>{\hspace{-1pt}>\hspace{-1pt}}
\def\<{\hspace{-1pt}<\hspace{-1pt}}
\def\ie{\textit{i.e.}}
\def\eg{\textit{e.g.} }

\newcommand{\msf}[1]{\mathsf{#1}}

\newcommand{\mrm}[1]{\mathrm{#1}}

\def\eqdef{\triangleq}

\newlength\ploth 
\newlength\plotw 

\pgfplotsset{
	dirac/.style={
	mark={triangle*},
	mark size = 1.5pt,
	ycomb,
	scatter,
	visualization depends on={(y/abs(y)-1)/2 \as \sign},
	scatter/@pre marker code/.code={\scope[rotate=180*\sign,yshift=-2pt]}
    }
}

\tikzset{
   precise pin/.style args={#1:#2:#3}{
        pin={[
            inner sep=0pt,
            pin distance=#3,
            label={[
                append after command={
                    node [
                        inner sep=-1pt,
                        at=(\tikzlastnode.#1),
                        anchor=#1+180
                    ] {#2}
                }
            ]center:{}}
        ]#1:{}}
    }
}

\tikzset{small dot/.style={fill=black, circle,scale=0.2}}

\let\Gamma\varGamma
\let\Delta\varDelta
\let\Theta\varTheta
\let\Lambda\varLambda
\let\Xi\varXi
\let\Pi\varPi
\let\Sigma\varSigma
\let\Upsilon\varUpsilon
\let\Phi\varPhi
\let\Psi\varPsi
\let\Omega\varOmega

\makeatletter
\def\grd@save@target#1{%
  \def\grd@target{#1}}
\def\grd@save@start#1{%
  \def\grd@start{#1}}
\tikzset{
  grid with coordinates/.style={
    to path={%
      \pgfextra{%
        \edef\grd@@target{(\tikztotarget)}%
        \tikz@scan@one@point\grd@save@target\grd@@target\relax
        \edef\grd@@start{(\tikztostart)}%
        \tikz@scan@one@point\grd@save@start\grd@@start\relax
        \draw[minor help lines] (\tikztostart) grid (\tikztotarget);
        \draw[major help lines] (\tikztostart) grid (\tikztotarget);
        \grd@start
        \pgfmathsetmacro{\grd@xa}{\the\pgf@x/1cm}
        \pgfmathsetmacro{\grd@ya}{\the\pgf@y/1cm}
        \grd@target
        \pgfmathsetmacro{\grd@xb}{\the\pgf@x/1cm}
        \pgfmathsetmacro{\grd@yb}{\the\pgf@y/1cm}
        \pgfmathsetmacro{\grd@xc}{\grd@xa + \pgfkeysvalueof{/tikz/grid with coordinates/major step}}
        \pgfmathsetmacro{\grd@yc}{\grd@ya + \pgfkeysvalueof{/tikz/grid with coordinates/major step}}
        \foreach \x in {\grd@xa,\grd@xc,...,\grd@xb}
        \node[anchor=north] at (\x,\grd@ya) {\pgfmathprintnumber{\x}};
        \foreach \y in {\grd@ya,\grd@yc,...,\grd@yb}
        \node[anchor=east] at (\grd@xa,\y) {\pgfmathprintnumber{\y}};
      }
    }
  },
  minor help lines/.style={
    help lines,
    step=\pgfkeysvalueof{/tikz/grid with coordinates/minor step}
  },
  major help lines/.style={
    help lines,
    line width=\pgfkeysvalueof{/tikz/grid with coordinates/major line width},
    step=\pgfkeysvalueof{/tikz/grid with coordinates/major step}
  },
  grid with coordinates/.cd,
  minor step/.initial=.1,
  major step/.initial=1,
  major line width/.initial=0.5pt,
}
\makeatother

%% file: sec_prerr.tex
\section{Probability of Error}\label{sec:poe}

\subsection{Single User}


The main contribution of this subsection is to show that imperfect TR and AR achieves the same probability of error, and, therefore, that the same accuracy is needed for channel estimation at transmitter and receiver in order to achieve a given error probability.

With reference to decision variables $\hat{z}^\mrm{TR}$ of eq.~\eqref{eq:zTR} and $\hat{z}^\mrm{AR}$ of eq.~\eqref{eq:decR}, the probability of error, in both cases, is:
\begin{align} \label{eq:PeAR}
 P_e 	 = \frac{1}{2}\Prob{z<0\;|\;b=\msf{A}}+\frac{1}{2}\Prob{z>0\;|\;b=-\msf{A}} 
 		 = \Prob{z<0\;|\;b=\msf{A}},
\end{align}
where the first equality follows from $b$ belonging to $\{-\msf{A},\msf{A}\}$ with equal probability, and the second equality follows from the distribution of $z$ being an even function. For the power constraint, it results $\msf{A}=\sqrt{\En}$. Equivalence of $P_e$ for the two cases is derived by showing that $\hat{z}^\mrm{TR}$ and $\hat{z}^\mrm{AR}$ have the same distribution. 

To this end, rewrite the decision variable $\hat{z}^\mrm{AR}$ conditioned on $b=\msf{A}$. Without loss of generality, and for the sake of simplicity, consider $\bs{x}=\bs{e}_1$. Then:
\begin{align*} \hat{z}^\mrm{AR} 	 = \|\bs{c}\|^2 \msf{A} + \bs{c}^\t (\bs{n}+\bs{\xi}\. \msf{A}) + \bs{\xi}^\t \bs{n} 
					 = \|\bs{c}\|^2 \msf{A} + \bs{c}^\t \bs{\xi}\. \msf{A} + \bs{n}^\t(\bs{\xi}+\bs{c}).
\end{align*}
Similarly, the decision variable $\hat{z}^\mrm{TR}$ conditioned on $b=\msf{A}$ is:
\[ \hat{z}^\mrm{TR} = \msf{A}\,\bs{c}^\t (\bs{c}+\bs{\xi}) + n\|\bs{c}+\bs{\xi}\|, \]
where $n\sim\mathcal{N}(0,\sigman^2)$. By comparing the two expressions, $\hat{z}^\mrm{AR}$ is equivalent to $\hat{z}^\mrm{TR}$, the equivalence being defined as producing the same $P_e$, iff term $\bs{n}^\t(\bs{\xi}+\bs{c})$ in $\hat{z}^\mrm{AR}$ is distributed as $n\|\bs{c}+\bs{\xi}\|$ in $\hat{z}^\mrm{TR}$. This is, indeed, the case; by choosing an orthogonal matrix $\bs{Q}$ such that $\bs{Q}(\bs{c}+\bs{\xi})=\| \bs{c}+\bs{\xi} \|\bs{e}_1$, one has:
\[ \bs{n}^\t(\bs{c}+\bs{\xi}) = \bs{n}^\t\bs{Q}^\t\bs{Q}(\bs{c}+\bs{\xi}) = \bs{n}_1'^\t \|\bs{c}+\bs{\xi}\| \bs{e}_1 = n_1' \|\bs{c}+\bs{\xi}\|, \]
where $n_1'\sim\mathcal{N}(0,\sigman^2)$, hence the equivalence in terms of distributions, and, therefore, probability of error is verified.

\subsection{Multiuser}

In the multiuser setting, although the expression for the probability of error remains as in eq.~\eqref{eq:PeAR}, there are three sources of errors: thermal noise, imperfect CSI, and multiuser interference (MUI). In particular, as $\En/\sigman^2$ increases, the last two factors both lead to a probability error floor, \ie, $P_e\to P_e^\textup{floor}(\beta,\sigma_\xi^2)>0$ as ${\En/\sigman^2\to\infty}$, being $\beta=K/N$ the load of the system. 

\begin{figure}[t]
\centering
\setlength\ploth{1.4\columnwidth} 
\setlength\plotw{0.9\columnwidth}
\subfigure[\emph{TR}]{ \label{fig:ber1}\includegraphics[scale=0.9]{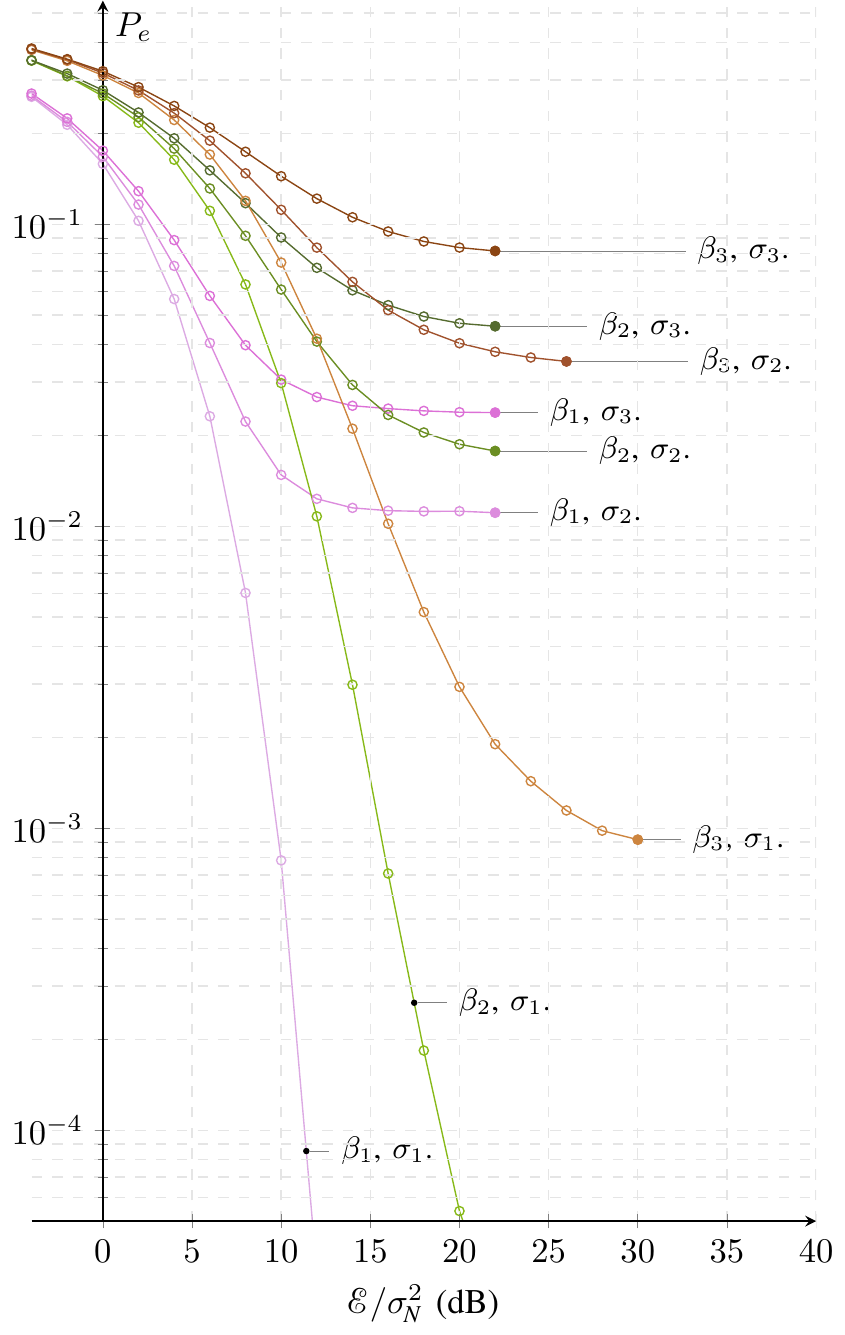} }
\hfill
\subfigure[\emph{AR}]{ \label{fig:ber2}\includegraphics[scale=0.9]{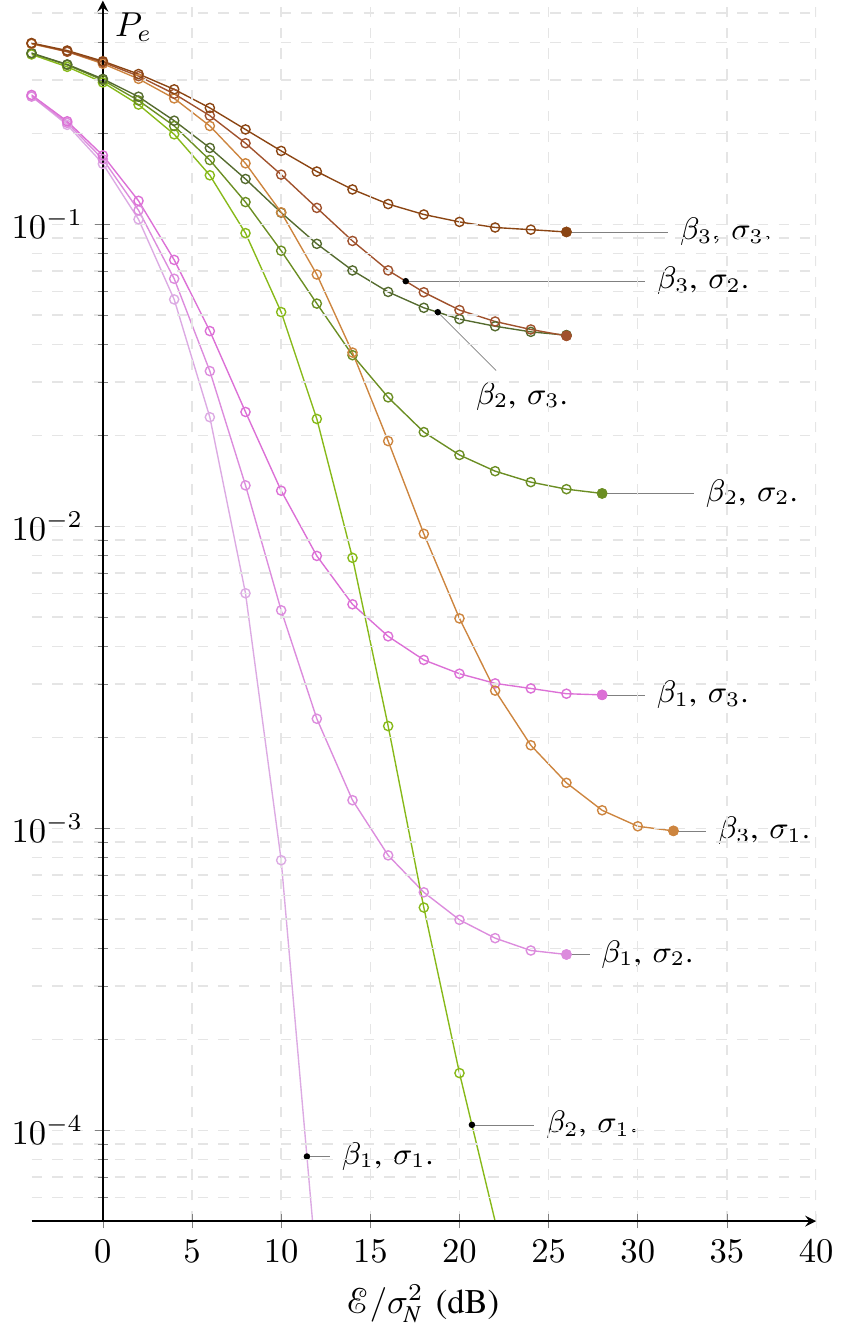} }
\caption{Probability of error $P_e$ vs. $\En/\sigman^2$ (dB) for systems with $\ii=1$ and $(\beta_1,\beta_2,\beta_3)=(\sigma_1^2,\sigma_2^2,\sigma_3^2)=(0,\frac{1}{20},\frac{1}{10})$. Figure \textit{(a)} refers to TR while \textit{(b)} to AR.}
\label{fig:BERnonimp}
\end{figure}

\begin{figure}[t]
\centering
\setlength\ploth{1.4\columnwidth} 
\setlength\plotw{0.9\columnwidth}
\subfigure[\emph{TR}]{ \label{fig:ber3}\includegraphics[scale=0.9]{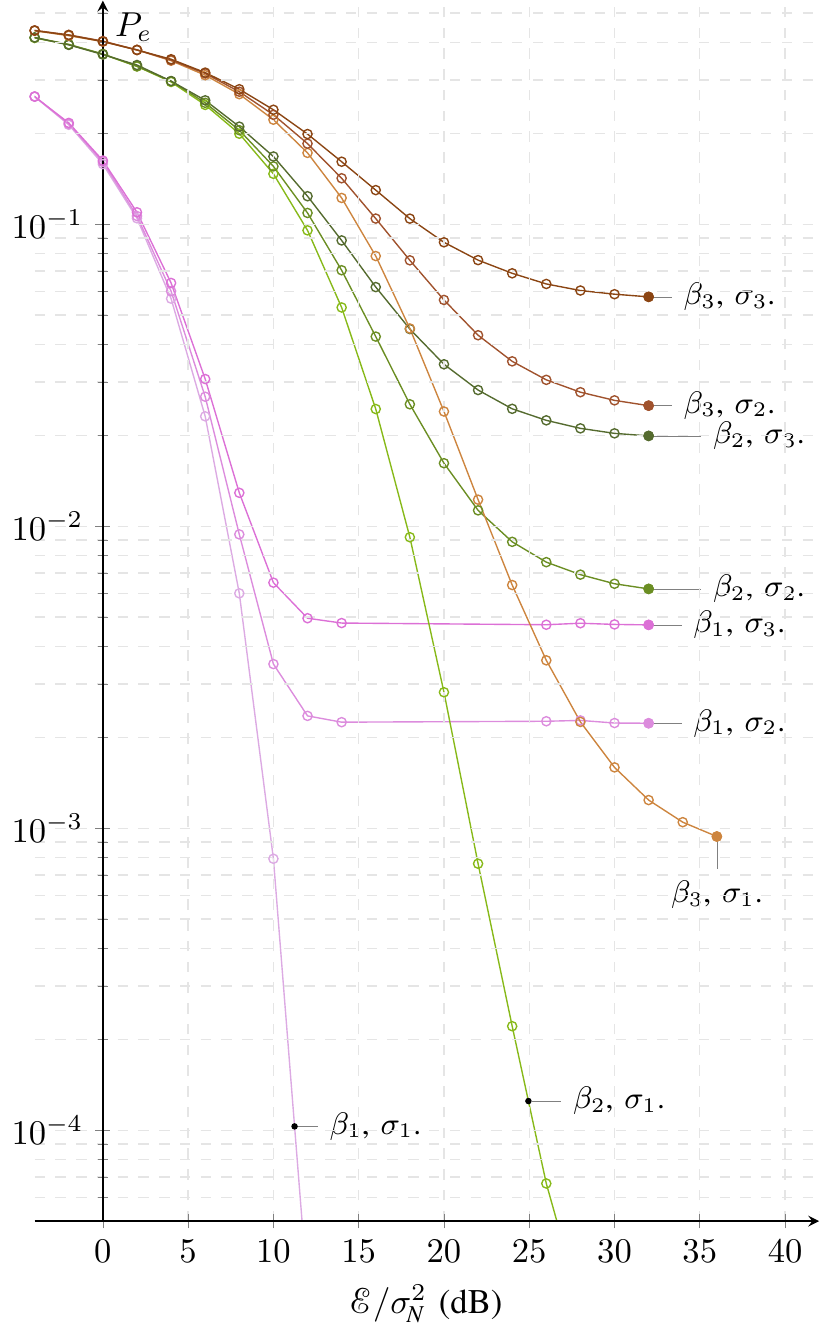}  }
\hfill
\subfigure[\emph{AR}]{ \label{fig:ber4}\includegraphics[scale=0.9]{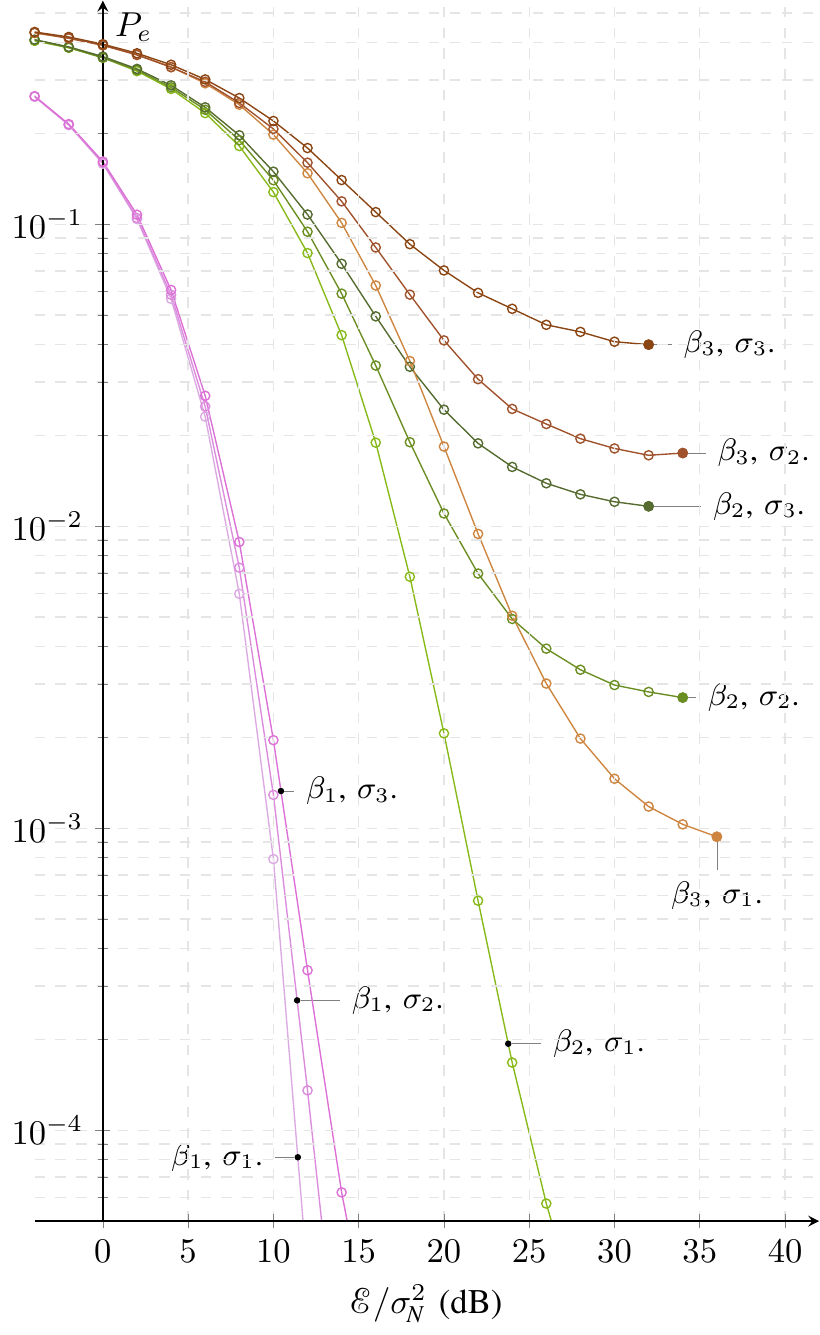}  }
\caption{Probability of error $P_e$ vs. $\En/\sigman^2$ (dB) for impulsive ($\ii=5$) systems with $(\beta_1,\beta_2,\beta_3)=(\sigma_1^2,\sigma_2^2,\sigma_3^2)=(0,\frac{1}{20},\frac{1}{10})$. Figure \textit{(a)} refers to TR while \textit{(b)} to AR.}
\label{fig:BERimp}
\end{figure}

Figures~\ref{fig:BERnonimp}~and~\ref{fig:BERimp} show the probability of error $P_e$ vs. $\En/\sigman^2$ (dB) for systems with $\ii=1$ vs. $\ii=5$, respectively, and for different values of $\beta$ and $\sigma_\xi^2$, $(\beta_1,\beta_2,\beta_3)=(\sigma_1^2,\sigma_2^2,\sigma_3^2)=(0,\frac{1}{20},\frac{1}{10})$. In particular, in both figures, the left-hand side plot (Fig.~\ref{fig:ber1} and~\ref{fig:ber3}) refers to TR, while the r.h.s. plot (Fig.~\ref{fig:ber2} and~\ref{fig:ber4}) refers to AR. Figures show that floors depend on a combination of both imperfect CSI and MUI. For low-SNR, \ie, $\En\ll\sigman^2$, $P_e$ is not very sensitive to estimation errors: on figures, systems with same load have similar $P_e$ at low-SNR, and leads to different $P_e^\textup{floor}$ at high-SNR, $\En\gg\sigman^2$. On the contrary, for fixed estimation error variance $\sigma_\xi^2$, increased $\beta$ implies increased $P_e$ for any SNR, and contributes to a higher error floor. All figures are obtained by Monte-Carlo simulations of finite-dimensional systems with $N=200$ chips. By comparing Figs.~\ref{fig:BERnonimp} \textit{(a)} and \textit{(b)}, that both refer to systems with $\ii=1$, AR is shown to outperform TR when imperfect CSI is the main cause of error, and vice versa when the load is the main cause of error: compare, for example, the two cases $(\beta_3,\sigma_1)$ and $(\beta_1,\sigma_2)$. A similar behavior can be observed with impulsive systems (Fig.~\ref{fig:BERimp}) with even more emphasis. Figure~\ref{fig:BERimp} \textit{(a)} and \textit{(b)} shows performance of systems with $\ii=5$, and indicates that  performance of TR is strongly limited by the presence of estimation errors, while when MUI is the limiting cause TR and AR have similar performance (see for example curves with $(\beta_3,\sigma_1)$ and $(\beta_1,\sigma_2)$).   

%% file: trantr_sc.bbl
\begin{thebibliography}{1}

%

%
%
\bibitem{biglieri:coding}
Biglieri E., ``Coding for Wireless Channels,'' \emph{Springer}, 2005.

\bibitem{caire:bestpaper}
Caire G., Jindal N., Kobayashi M., Ravindran N., ``Multiuser MIMO Achievable Rates With Downlink Training and Channel State Feedback,'' \emph{IEEE Transactions on Information Theory}, Vol. 56, No. 6, Jun 2010. 


\bibitem{chen:math}
Chen~K-M., ``A Mathematical Formulation of the Equivalence Principle,'' \emph{IEEE Transactions on Antennas Propagat.}, AP-37, 1576, 1989.

\bibitem{denardis:uwbtr}
De Nardis L., Fiorina J., Panaitopol D., Di Benedetto M.-G., ``Combining UWB with Time Reversal for improved communication and positioning,'' \emph{Telecommunication System Journal (Springer), Special Issue on Recent Advances In UWB Systems: Theory and Applications}, 2011, pp. 1-14.

\bibitem{derode:aps}
Derode A., Roux P., Fink M.,``Robust acoustic time reversal with high-order multiple scattering'', \emph{Phys. Rev. Letters}, 1995, pp. 4206-4209, Vol. 75.

\bibitem{MGDB:book}
Di Benedetto M.-G., Giancola G., ``{Understanding Ultra Wide Band, Radio Fundamentals},'' \emph{Prentice Hall Pearson Education}, 2004.

\bibitem{esma:rake1}
Esmailzadeh R., Nakagawa M., ``Pre-RAKE Diversity Combination for Direct Sequence Spread Spectrum Communications Systems,'' \emph{IEEE ICC '93}, Geneva, pp. 463-467, Vol. 1, 1993.

\bibitem{esma:rake2} 
Esmailzadeh R., Nakagawa M., ``Pre-RAKE Diversity Combination for Direct Sequence Spread Spectrum Mobile Communications Systems,'' \emph{IEICE Transactions on Communications}, Vol. E76-B, No. 8, pp. 1008Ð1015, Aug. 1993.
%
%

%
%

%
%
%
%
%
\bibitem{guido:poznan12}
Ferrante G.~C., ``Time Reversal Against Optimum Precoder Over Frequency-Selective Channels,'' \emph{European Wireless 2012,} Poznan, Poland, Apr 18-20, 2012.

\bibitem{ferr:wcnc}
Ferrante G. C., Fiorina J., Di Benedetto M.-G., ``Complexity reduction by combining time reversal and IR-UWB,'' \emph{2012 IEEE Wireless Communications and Networking Conference (WCNC)}, pp. 28,31, 1-4 Apr. 2012.

\bibitem{ferr:icuwb13}
Ferrante G. C., Fiorina J., Di Benedetto M.-G., ``Time Reversal Beamforming in MISO-UWB Channels,'' \emph{2013 IEEE International Conference on Ultra-Wideband (ICUWB)}, Sydney, Sep. 15-18, 2013.

\bibitem{fiorina:icuwb11}
Fiorina, J., Capodanno G., Di Benedetto M.-G., ``Impact of Time Reversal on multi-user interference in IR-UWB,'' \emph{2011 IEEE International Conference on Ultra-Wideband (ICUWB)}, pp. 415-419, Sep. 14-16, 2011.

\bibitem{fink:old}
Fink M., ``Time Reversal of Ultrasonic Fields---Part I: Basic Principles,'' \emph{IEEE Transactions on Ultrason., Ferroelectr. Freq. Control}, Vol. 39, No. 5, pp.
555-566, 1992.

\bibitem{fink:acoustic}
Fink M., Montaldo G., Tanter M., `Time Reversal Acoustics,'' \emph{IEEE International Ultrasonics, Ferroelectrics, and Frequency Control Joint 50th Anniversary Conference,} pp. 850-859, 2004.

\bibitem{foerster:chmodel}
Foerster J., \emph{Channel Modeling Sub-Committee report final}, IEEE Doc. P802.15-02/490r1-SG3a, 2003.


\bibitem{hassibihochwald:training}
Hassibi B., Hochwald B.~M., ``How Much Training is Needed in Multiple-Antenna Wireless Links?,'' \emph{IEEE Transactions on Information Theory,} Vol. 49, No. 4, pp. 951-963, Apr. 2003.

\bibitem{joham:equiv}
Joham M., Utschick W., Nossek~J.~A., ``On the Equivalence of Prerake and Transmit Matched Filter,'' \emph{Proc. ASST 2001}, pp. 313Ð318, Sep. 2001.

\bibitem{joham:mmse}
Joham M., Kusume K., Gzara M. H., Utschick W., Nossek J. A., ``Transmit Wiener Filter for the Downlink of TDD DS-CDMA Systems,'' \emph{IEEE 7th Int. Symp. on Spread-Spectrum Tech. and Appli.}, Prague, Sep. 2-5 2002.

\bibitem{lapidoth:howmuch}
Lapidoth~A., Shamai S., ``Fading Channels: How Perfect Need 'Perfect Side Information' Need?,'' \emph{IEEE Transactions on Information Theory,} Vol. 48, No. 5, pp. 1118-1134, 2002.

\bibitem{Larsson:mimo}
Larsson E.G., Stoica P., ``Space-Time Block Coding for Wireless Communications,'' \emph{Cambridge University Press}, 2003.
%
\bibitem{merhav:rates}
Merhav N., Kaplan G., Lapidoth A., Shitz S.~S., ``On information rates for mismatched decoders,'' \emph{IEEE Transactions on Information Theory}, Vol. 40, pp. 1953Ð1967, Nov. 1994.

\bibitem{mitra:cross}
Mitra A., ``On Pseudo-Random and Orthogonal Binary Spreading Sequences,'' \emph{International Journal of Information and Communication Engineering,} 4:6, pp. 447-454, 2008.

\bibitem{ngu:trmimo}
Nguyen H. T., Andersen J. B., Pedersen G. F., Kyritsi P., Eggers P. C. F., ``Time Reversal in Wireless Communications: A Measurement-Based Investigation,'' \emph{IEEE Transactions on Wireless Communications}, Vol. 5, No. 8, pp. 2242-2252, Aug. 2006.

\bibitem{noll:ss}
Noll Barreto A., Fettweis G., ``Capacity Increase in the Downlink of Spread Spectrum Systems through Joint Signal Precoding,'' \emph{Proc. ICC 2001}, Vol. 4, pp. 1142-1146, Jun. 2001.
%
%
\bibitem{palo:mono}
Palomar D. P., Jiang Y., ``MIMO Transceiver Design via Majorization Theory,'' \emph{Foundation and Trends in Communications and Information Theory,} Vol. 3, No.s 4-5, pp. 331-551, 2006.


\bibitem{pursley:cdma}
Pursley M. B., ``Performance Evaluation for Phase-Coded Spread-Spectrum Multiple-Access Communication---Part I: System Analysis,'' IEEE Transactions on Communications, Vol. 25, No. 8, Aug. 1977.
\bibitem{qiu:symp}
Akogun A.E., Qiu R.C., Guo N., ``Demonstrating Time Reversal in Ultra-wideband Communications Using Time Domain Measurements,'' \emph{51st International Instrumentation Symposium}, Knoxville, Tennessee, 8-12 May 2005.
\bibitem{sarwate:cross}
Sarwate D.~V., Pursley M.~B., ``Crosscorrelation Properties of Pseudorandom and Related Sequences,'' \emph{Proceedings of the IEEE,} Vol. 68, No. 5, May 1980.
\bibitem{shelk:uno}
Schelkunoff~S.~A., ``Some Equivalent Theorems of Electromagnetics and their Application to Radiation Problems,'' \emph{Bell Sys. Tech. J.}, 15, 92, 1936.
\bibitem{shelk:due}
Schelkunoff~S.~A., ``On Diffraction and Radiation of Electromagnetic Waves,'' \emph{Phys. rev.}, 56, 308, 1939.
\bibitem{stratton:uno}
Stratton~J.~A., Chu~L.~J., ``Diffraction Theory of Electromagnetic Waves,'' \emph{Phys. Rev.}, 56, 99, 1939. 
%
%
%
\bibitem{tang:zfds} 
Tang~Z., Cheng S., ``Interference Cancellation for DS-CDMA Systems over Flat Fading Channels through Pre-decorrelating,'' \emph{Proc. PIMRC '94}, Vol. II, pp. 435Ð438, Sep. 1994.
\bibitem{precoding:overview}
Vu M., Paulraj A., ``MIMO Wireless Linear Precoding,'' \emph{IEEE Signal Processing Magazine,} pp. 86-105, Sep. 2007.
\bibitem{qiu:misoexp1}
Qiu R., Zhou C., Guo N., Zhang J.~Q., ``Time Reversal With MISO for Ultra-wideband Communications: Experimental Results,'' \emph{IEEE Antennas and Wireless Propagation Letters,} Vol. 5, pp. 269-273, 2006.
\bibitem{qiu:misoexp2}
Guo N., Sadler B.~M., Qiu R., ``Reduced-Complexity UWB Time-Reversal Techniques and Experimental Results,'' \emph{IEEE Transactions on Wireless Communications,} Vol. 6, No. 12, pp. 4221-4226, 2007.
\bibitem{qiu:gen1}
Qiu R.~C., Zhang J. Q, Guo N., ``Detection of Physics-Based Ultra-Wideband Signals Using Generalized RAKE With Multiuser Detection (MUD) and Time-Reversal Mirror,'' \emph{IEEE Journal on Selected Areas in Communications,} Vol. 24, No. 4, pp. 724-730, 2006.
\bibitem{qiu:gen2}
Qiu R.~C., ``A Generalized Time Domain Multipath Channel and its Application in Ultra-Wideband (UWB) Wireless Optimal Receiver --- Part III: System Performance Analysis,'' \emph{IEEE Transactions on Wireless Communications,} Vol. 5, No. 10, pp. 2685-2695, 2006.
\bibitem{qiu:mimo}
Zhou C., Guo N., Qiu R. C., ``Time-Reversed Ultra-wideband (UWB) Multiple Input Multiple Output (MIMO) Based on Measured Spatial Channels,'' \emph{IEEE Transactions on Vehicular Technology,} Vol. 58, No. 6, pp. 2884-2898, 2009.
\bibitem{stro:tr}
Strohmer T., Emami M., Hansen J., Pananicolaou G., Paulraj A. J., ``Application of Time-Reversal with MMSE Equalizer to UWB Communications,'' \emph{Proc. Globecom '04}, pp. 3123-3127, Dec 2004.
\bibitem{verdu:mud}
Verd\'u S., ``Multiuser detection,'' \emph{Cambridge University Press}, 1998.
\bibitem{voj:tx} 
Voj\u{c}i\'{c} B. R., Jang W. M., ``Transmitter Precoding in Synchronous Multiuser Communications,'' \emph{IEEE Transactions on Communications}, pp. 1346Ð1355, Vol. 46, No. 10, Oct 1998.
\bibitem{sens:cdma}
Wang X., Wang J., ``Effect of Imperfect Channel Estimation on Transmit Diversity in CDMA Systems,'' \emph{IEEE Transactions on Vehicular Technology,} Vol. 53, No. 5, pp. 1400-1412, 2004.
\bibitem{win:perfrake}
Win M.~Z., Chrisikos G., Sollenberger N.R., `Performance of Rake Reception in Dense Multipath Channels: Implications of Spreading Bandwidth and Selection Diversity Order,'' \emph{IEEE Journal on Selected Areas in Communications}, Vol. 18, No. 8, pp. 1516-1525, Aug. 2000.
\bibitem{welch:bounds}
Welch L.~R., ``Lower Bounds on the Maximum Cross Correlation of Signals,'' \emph{IEEE Transactions on Information Theory}, pp. 397-399, May 1974.

\end{thebibliography}
